\documentclass[a4paper,11pt]{article}

\usepackage{a4wide}
\usepackage[english]{babel}
\usepackage{amsmath,amssymb,amsfonts,amsthm}
\usepackage{color}
\usepackage{graphicx}
\usepackage[utf8]{inputenc}
\usepackage{dsfont}
\usepackage{collref}
\usepackage{booktabs}
\usepackage{slashed}

\usepackage{hyperref}
\definecolor{darkred}{rgb}{0.8,0.1,0.1}
\hypersetup{
    colorlinks=true,         
    linkcolor=darkred, 
    citecolor=blue, 
}

\newcommand{\N}[1]{\ensuremath{N{=}#1}}

\newcommand{\calR}{\mathcal R}
\newcommand{\calo}{\ensuremath{o}}
\newcommand{\Hterm}[3]{\Big(H_{#1}^{#2} #3 h_{#1}^{#2}\Big)}
\newcommand{\Dleftslash}{ \stackrel{\leftarrow}{\rule[1.3ex]{0pt}{0pt}\smash[t]{\slashed{D}}} }

\DeclareMathOperator{\tr}{tr}
\newcommand{\sbr}[1]{{\scriptscriptstyle (#1)}}

\usepackage{mathrsfs}

\title{The Boundary~Multiplet~of
   \N{4}~SU(2)$\otimes$U(1)~Gauged~Supergravity
   on~Asymptotically-AdS$_5$}
\author{
  Thorsten Ohl\thanks{e-mail: \texttt{ohl@physik.uni-wuerzburg.de}}\qquad
  Christoph F.~Uhlemann\thanks{e-mail: \texttt{uhlemann@physik.uni-wuerzburg.de}} \\
  \hfill\\
  Institut f\"ur Theoretische Physik und Astrophysik\\
    Universit\"at W\"urzburg\\
  Am Hubland, 97074 W\"urzburg, Germany}

\begin{document}
\maketitle
\begin{abstract}
We consider \N{4} SU(2)$\otimes$U(1) gauged supergravity on asymptotically-AdS$_5$
backgrounds.  
By a near-boundary analysis we determine the boundary-dominant
components of the bulk fields from their partially gauge-fixed field equations.
Subdominant components are projected out in the boundary limit and we find a 
reduced set of boundary fields, constituting the \N{2} Weyl multiplet.
The residual bulk symmetries are found to act on the boundary fields as
four-dimensional diffeomorphisms, \N{2} supersymmetry and (super-)Weyl transformations. 
This shows that
the on-shell \N{4} supergravity multiplet yields the \N{2} Weyl multiplet on the boundary
with the appropriate local \N{2} superconformal transformations.
Building on these results we use the AdS/CFT conjecture to calculate the
Weyl anomaly of the dual four-dimensional superconformal field theories in a generic bosonic 
\N{2} conformal supergravity background.
\end{abstract}

\section{Introduction}
Supergravities on Anti-de Sitter (AdS) spaces play a prominent role in the
AdS/CFT correspondence \cite{Maldacena:1997re}, 
which -- in the weakest form of the conjecture -- relates
classical ten-dimensional supergravity on the near-horizon limit of 
$p$-brane backgrounds to strongly-coupled 
superconformal quantum field theories (SCFT) on $p{+}1$-dimensional flat space.
The near-horizon geometry of the $p$-brane solutions is typically given by a product of AdS space and a compact manifold, 
on which one may perform a Kaluza-Klein expansion.
Gauged supergravities on AdS spaces are then employed to describe the Kaluza-Klein expanded ten-dimensional 
theory truncated to a finite number of Kaluza-Klein modes,
and consequently also for a dual description of the 
corresponding SCFT sector \cite{Cvetic:2000nc,Ferrara:1998ej,Ferrara:1998ur}.
The explicit AdS/CFT duality relation is given by interpreting the boundary values of the supergravity fields 
as sources for the dual operators of the SCFT \cite{Gubser:1998bc,Witten:1998qj}, 
and it has been applied to describe a variety of phenomena in strongly-coupled 
Quantum Field Theories (QFT) \cite{Erdmenger:2007cm,Herzog:2009xv}.

In this work we consider five-dimensional half-maximally supersymmetric gauged supergravity.
The general gauged matter-coupled \N{4} supergravities in five dimensions were constructed 
in \cite{Dall'Agata:2001vb, Schon:2006kz}, and
it was noted in \cite{Dall'Agata:2001vb} that AdS ground states are only possible if the gauge group is a product
of a one-dimensional Abelian factor and a semi-simple group.
We focus on the \N{4} SU(2)$\otimes$U(1) gauged supergravity constructed by Romans \cite{Romans:1985ps}, 
the only gauging of the pure supergravity without additional matter multiplets which admits an AdS vacuum. 
Solutions of this theory can be lifted to solutions of the IIB supergravity \cite{Lu:1999bw} where they correspond 
to product geometries involving $S^5$, 
and also to warped-product solutions of IIA supergravity 
and the maximal $d{=}11$ supergravity \cite{Cvetic:2000yp,Gauntlett:2007sm}.
We restrict the configuration space to asymptotically-AdS$_5$ geometries
with an arbitrary four-dimensional boundary metric.
By an analysis of the asymptotic field equations we determine the multiplet of boundary fields, 
and from the local bulk symmetries
we obtain the boundary symmetries with the induced representation on the boundary fields.
This limiting procedure does not involve the AdS/CFT conjecture and does not rely on the choice of boundary conditions.
We find the \N{2} Weyl multiplet with local \N{2} superconformal transformations.
Similar calculations have previously been carried out for bulk theories
in $d{=}3,6,7$ dimensions and for \N{2} supergravity in $d{=}5$ 
\cite{Nishimura:1998ud,Nishimura:1999gg,Nishimura:1999av,Nishimura:2000wj,Banados:1998pi,Henneaux:1999ib,Balasubramanian:2000pq}.
Having established the asymptotic behaviour of the bulk fields and their symmetry transformations
we then 
present a first application using the AdS/CFT conjecture.
For the bosonic sector of the bulk supergravity
we carry out the 
holographic renormalization \cite{Henningson:1998gx,Balasubramanian:1999re,deHaro:2000xn}
and
calculate the Weyl anomaly of the dual four-dimensional SCFTs
in a generic bosonic \N{2} conformal supergravity background.
This extends the existing results for nontrivial metric and dilaton 
backgrounds \cite{Henningson:1998gx,Nojiri:1998dh,Anselmi:1998zb,Nojiri:1999mh,Blau:1999vz}\footnote{%
For the maximally supersymmetric case a discussion of the SCFT effective action, the conformal anomaly and the role of 
conformal supergravity in AdS/CFT can be found in \cite{Liu:1998bu}.
Explicit constructions for the boundary of AdS are given there for the metric-dilaton sector.
}.

Our results on the asymptotic structure of the \N{4} gauged supergravity 
may also be relevant in the following context.
A duality relation of QFTs on AdS space and on 
its conformal boundary has been formulated and proven in \cite{Rehren:1999jn, Rehren:2000tp} 
in the framework of algebraic QFT. 
In contrast to the AdS/CFT correspondence,
gravity does not seem to play a dedicated role in the algebraic holography.
In particular, the constructions in \cite{Bertola:2000pp, Rehren:2004yu} suggest that a gravitational theory is induced 
on the conformal boundary by a gravitational bulk theory.
A similar result was obtained in \cite{Compere:2008us} by deforming the AdS/CFT correspondence.
It was shown there that changing the Dirichlet boundary conditions to Neumann or mixed boundary conditions 
promotes the boundary metric to a dynamical field.
In this context our construction yields the kinematics of the boundary theory, 
for which we thus expect an \N{2} conformal supergravity. 

The paper is organized as follows. 
In Section~\ref{sec:RomansN4} we review the \N{4} SU(2)$\otimes$U(1) 
gauged supergravity \cite{Romans:1985ps} to fix notation.
In Section~\ref{sec:boundary-limit} the notion of an asymptotically-AdS$_5$ space is introduced
and the multiplet of fields induced on the conformal boundary  is constructed.
We employ Fefferman-Graham coordinates and partial gauge fixing of the local super-, 
Lorentz and SU(2)$\otimes$U(1) symmetries.
The asymptotic scalings of the boundary-irreducible components of the bulk fields are determined 
in Section~\ref{sec:boundary-fields} from the linearized field equations.
Subdominant components are projected out in the boundary limit and we find a reduced set 
of boundary fields, constituting the \N{2} Weyl multiplet.
These results are extended to the nonlinear theory in Section~\ref{sec:nonlinear-theory}, where
we argue for the consistency of the previous construction with the interaction terms. 
We also determine those of the subdominant bulk field components which then enter the boundary symmetry transformations.
The residual bulk symmetries preserving the gauge fixings, and their action on the boundary fields
are determined in Section~\ref{sec:boundary-symmetries}. 
This yields the complete local \N{2} superconformal transformations of the Weyl multiplet.
In Section~\ref{sec:Weyl-anomaly} we use the AdS/CFT correspondence to calculate the Weyl anomaly of the dual SCFTs
in an external bosonic \N{2} conformal supergravity background.
To this end we determine the required subleading modes of the bulk fields in Section~\ref{sec:Weyl-anomaly-on-shell-fields}
and carry out the holographic renormalization in Section~\ref{sec:Weyl-anomaly-holographic-renormalization}.
We conclude in Section~\ref{sec:conclusions}.
Two appendices contain an overview of our conventions and connect the results of Section \ref{sec:boundary-symmetries} 
to the literature on \N{2} supergravity multiplets.

\section{Romans' \texorpdfstring{$\mathbf{\N{4}}$}{N=4} SU(2)\texorpdfstring{$\mathbf{\otimes}$}{x}U(1) gauged supergravity} \label{sec:RomansN4}
In this section we briefly discuss the five-dimensional gauged supergravity 
\cite{Romans:1985ps} in order to fix notation.
The theory has \N{4} supersymmetry (counted in terms of symplectic Majorana spinors) with $R$-symmetry group USp(4),
of which an SU(2)$\otimes$U(1) subgroup is gauged.
The symplectic metric is denoted by $\Omega$, and 
exploiting the isomorphism 
$\mathfrak{usp}(4)\cong\mathfrak{so}(5)$
the Lie algebra generators are given by
$\Gamma_{mn}:=\frac{1}{2}\left[\Gamma_m,\Gamma_n\right]$ 
with $\mathfrak{so}(5)$ vector indices $m,n$, and $\Gamma_m$ satisfying 
the five-dimensional Euclidean Clifford algebra 
relation\footnote{\label{usp4isomorphism}%
The $\Gamma_m$ can all be chosen hermitian, such that $\Gamma_{mn}^\dagger+\Gamma_{mn}=0$. 
With the charge conjugation matrix $C_E$ satisfying $C_E\Gamma_m C_E^{-1}=\Gamma_m^T$, we can identify 
$\Omega:=C_E$
and have $\Omega\Gamma_{mn}+\Gamma_{mn}^T \Omega=0$, 
providing the isomorphism $\mathfrak{usp}(4)\cong\mathfrak{so}(5)$.
} 
$\lbrace \Gamma_m,\Gamma_n\rbrace=2 \delta_{mn} \mathds{1}$.
With the obvious embedding of 
$\mathfrak{su}(2){\oplus}\mathfrak{u}(1)\cong\mathfrak{so}(3){\oplus}\mathfrak{so}(2)$
into $\mathfrak{usp}(4)\cong\mathfrak{so}(5)$, 
the vector index $m$ decomposes into $m=(I,\alpha)$ with $I=1,2,3$ and $\alpha=4,5$.
We consider the theory referred to
as \N{4^+} in \cite{Romans:1985ps}, for which the SU(2) gauge coupling $g_2$ 
is fixed in terms of the U(1) coupling $g_1$ by $g_2=+\sqrt{2}g_1=:g$. 
For this choice of couplings the theory admits an AdS solution.
The bosonic field content is given by the vielbein $e_\mu^a$, 
two antisymmetric tensor fields $B_{\mu\nu}^\alpha$,
the SU(2) and U(1) gauge fields $A_\mu^I$ and $a_\mu$, respectively, 
and a scalar $\varphi$.
The four gravitinos $\psi_\mu^i$ and four spin-$\frac{1}{2}$ fermions $\chi^i$
comprising the fermionic field content
are in the spinor $\mathbf{4}$ of $\mathfrak{usp}(4)$, which decomposes as 
$\mathbf{4}\rightarrow \mathbf{2}_{1/2} + \mathbf{2}_{-1/2}$. 
The vector and tensor fields originate from the vector representation, decomposing
as $\mathbf{5}\rightarrow \mathbf{3}_0+\mathbf{1}_1+\mathbf{1}_{-1}$.
The spinors satisfy the symplectic Majorana condition, e.g.~$\bar\chi^i=\left(\chi^i\right)^T C$ 
with the conjugate $\bar\chi^i:=\left(\chi_i\right)^\dagger\gamma_0$,
the metric is of signature $({+},{-},{-},{-},{-})$ and the $\gamma$-matrices are chosen such that
$\gamma_{abcde}=\epsilon_{abcde}$ with $\epsilon_{01234}=1$. 
For a summary of the conventions see Appendix \ref{app:conventions}.
From this point on we denote five-dimensional objects with hat 
and four-dimensional ones without, e.g.~five-dimensional spacetime indices $\hat\mu=(\mu,r)$ with $\mu=0,1,2,3$.
The Lagrangian as given up to four-fermion terms in \cite{Romans:1985ps} is
{\allowdisplaybreaks
\begin{align}\label{eqn:N4-Lagrangian1}
 \mathcal L=&-\frac{1}{4} \hat e\hat\calR(\hat\omega)
-\frac{1}{2}i\hat e\hat{\bar{\psi}}_{\hat\mu}^i\hat\gamma^{\hat\mu\hat\nu\hat\rho}\hat{D}_{\hat\nu}\hat\psi_{\hat\rho i}
+\frac{3}{2}i\hat e T_{ij}\hat{\bar{\psi}}_{\hat\mu}^{i}\hat\gamma^{\hat\mu\hat\nu}\hat \psi_{\hat\nu}^j
-i\hat e A_{ij}\hat{\bar\psi}_{\hat\mu}^i\hat\gamma^{\hat\mu}\hat\chi^j
+\frac{1}{2}i\hat e\hat{\bar\chi}^i\hat\gamma^{\hat\mu} \hat D_{\hat\mu}\hat\chi_i\nonumber\\
&
+i\hat e \Big(\frac{1}{2}T_{ij}-\frac{1}{\sqrt{3}}A_{ij}\Big)\hat{\bar\chi}^i\hat\chi^j
+\frac{1}{2}\hat e\hat D^{\hat\mu}\hat\varphi \hat D_{\hat\mu}\hat\varphi+\hat e P(\hat\varphi)
-\frac{1}{4}\hat e\,\xi^2 \hat B^{\hat\mu\hat\nu\alpha}\hat B_{\hat\mu\hat\nu}^{\ \alpha}\nonumber\\
&
+\frac{1}{4 g_1}\hat\epsilon^{\hat\mu\hat\nu\hat\rho\hat\sigma\hat\tau}\epsilon_{\alpha\beta}\hat B_{\hat\mu\hat\nu}^{\ \alpha}\hat D_{\hat\rho}\hat B_{\hat\sigma\hat\tau}^{\ \beta}
-\frac{1}{4}\hat e\,\xi^{-4}\hat f^{\hat\mu\hat\nu}\hat f_{\hat\mu\hat\nu}
-\frac{1}{4}\hat e\,\xi^2\hat F^{\hat\mu\hat\nu I} \hat F_{\hat\mu\hat\nu}^{\ I}
-\frac{1}{4}\hat\epsilon^{\hat\mu\hat\nu\hat\rho\hat\sigma\hat\tau}\hat F_{\hat\mu\hat\nu}^{\ I} \hat F_{\hat\rho\hat\sigma}^{\ I} \hat a_{\hat\tau}\\
&+\frac{1}{4\sqrt{2}}i \hat e \Hterm{\hat\mu\hat\nu}{ij}{+\frac{1}{\sqrt{2}}} \hat{\bar{\psi}}_i^{\hat\rho}\hat\gamma_{[\hat\rho}\hat\gamma^{\hat\mu\hat\nu}\hat\gamma_{\hat\sigma]}\hat\psi_j^{\hat\sigma}
 +\frac{1}{2\sqrt{6}}i \hat e\Hterm{\hat\mu\hat\nu}{ij}{-\sqrt{2}}\hat{\bar{\psi}}_i^{\hat\rho}\hat{\gamma}^{\hat\mu\hat\nu}\hat\gamma_{\hat\rho}\hat\chi_j \nonumber\\
&-\frac{1}{12\sqrt{2}}i \hat e\Hterm{\hat\mu\hat\nu}{ij}{-\frac{5}{\sqrt{2}}}\hat{\bar{\chi}}_i\hat\gamma^{\hat\mu\hat\nu}\hat\chi_j
 +\frac{1}{\sqrt{2}}i  \hat e\left(\partial_{\hat\nu}\hat\varphi\right)\hat{\bar{\psi}}_{\hat\mu}^i\hat\gamma^{\hat\nu}\hat\gamma^{\hat\mu}\hat\chi_i\nonumber
\end{align}
}%
with $\xi:=\operatorname{exp}{\sqrt{\frac{2}{3}}\hat\varphi}$ and the scalar potential $P(\hat\varphi):=\frac{1}{8}g^2\left(\xi^{-2}+2\xi\right)$. 
Antisymmetrization of indices is defined as $X_{[\mu}Y_{\nu]}:=\frac{1}{2}(X_\mu Y_\nu-X_\nu Y_\mu)$.
Furthermore, 
\begin{align}\label{eqn:defTij}
\begin{split}
T^{ij}:=\frac{g}{12\sqrt{2}}\left(2\xi^{-1}+\xi^2\right) \left(\Gamma_{45}\right)^{ij},&\qquad
A^{ij}:=\frac{g}{2\sqrt{6}}\left(\xi^{-1}-\xi^2\right)\left(\Gamma_{45}\right)^{ij}~,\\
H_{\hat\mu\hat\nu}^{ij}:=\xi\left(\hat F^I_{\hat\mu\hat\nu}\left(\Gamma_I\right)^{ij}+\hat B_{\hat\mu\hat\nu}^{\alpha}\left(\Gamma_{\alpha}\right)^{ij}\right)~,&\qquad
h_{\hat\mu\hat\nu}^{ij}:=\xi^{-2}\Omega^{ij}\hat f_{\hat\mu\hat\nu}~.
\end{split}
\end{align}
The covariant derivative on the spinor $\mathbf{4}$ of $\mathfrak{usp}(4)$ is given by
\begin{align}\label{eqn:covariant-derivative-spinor4}
 \hat D_{\hat\mu} v_i=\hat\nabla_{\hat\mu} v_i+\frac{1}{2}g_1 \hat a_{\hat\mu}\left(\Gamma_{45}\right)_i^{\ j} v_j+\frac{1}{2}g_2 \hat A_{\hat\mu}^I\left(\Gamma_{I45}\right)_i^{\ j}v_j~,
\end{align}
with the spacetime-covariant derivative $\hat\nabla_{\hat\mu}$
and
$\Gamma_{IJ}=-\epsilon^{IJK}\Gamma_{K45}$. 
Acting on a spinor $\hat\nabla_{\hat\mu}=\partial_{\hat\mu}+\frac{1}{4}\hat\omega_{\hat\mu}^{\hphantom{\hat\mu}\hat a\hat b}\hat\gamma_{\hat a\hat b}$, 
and
the curvatures are defined by
\begin{align}
 \big[\hat D_{\hat\mu},\hat D_{\hat\nu}\big]\hat\epsilon_i
  =:
 \frac{1}{4}\hat \calR_{\hat \mu\hat \nu}^{\hat a\hat b}(\hat\omega)\,\hat\gamma_{\hat a\hat b}\,\hat\epsilon_i
 +\frac{1}{2}g_1\hat f_{\hat\mu\hat\nu}\left(\Gamma_{45}\right)_i^{\ \,j}\hat\epsilon_j
 +\frac{1}{2}g_2\hat F^I_{\hat\mu\hat\nu}\left(\Gamma_{I45}\right)_i^{\ \,j}\hat\epsilon_j~.
\end{align}
On the vector $\mathbf{5}$ of $\mathfrak{usp}(4)$ the covariant derivative is given by
\begin{align}
\hat D_{\hat\mu} v^{I\alpha}=\hat\nabla_{\hat\mu} v^{I\alpha}+g_1 \hat a_{\hat\mu}\epsilon^{\alpha\beta}v^{I\beta}+g_2\epsilon^{IJK}\hat A_{\hat\mu}^Jv^{K\alpha}~.
\end{align}

The supersymmetry transformations to leading order in the fermionic terms are 
\begin{align}\label{N4-susy-variations}
\begin{split}
\delta_{\hat\epsilon} \hat e_{\hat\mu}^{\hat a}&=i\hat{\bar\psi}_{\hat\mu}^i\hat\gamma^{\hat a}\hat\varepsilon_i~,
\qquad
\delta_{\hat\epsilon} \hat A_{\hat \mu}^I=\Theta_{\hat\mu}^{ij}\left(\Gamma^I\right)_{ij},
\qquad
\delta_{\hat\epsilon} \hat\varphi=\frac{1}{\sqrt{2}}i\hat{\bar\chi}^i\hat\varepsilon_i~,
\\
\delta_{\hat\epsilon} \hat\psi_{\hat\mu i}&=\hat D_{\hat \mu}\hat\varepsilon_i
  +\hat\gamma_{\hat\mu}T_{ij}\hat\varepsilon^j
  -\frac{1}{6\sqrt{2}}\left(\hat\gamma_{\hat\mu}^{\ \:\hat\nu\hat\rho}  -4\delta_{\hat\mu}^{\ \:\hat\nu}\hat\gamma^{\hat\rho}\right)
  \Hterm{\hat\nu\hat\rho ij}{}{+\frac{1}{\sqrt{2}}}\hat\varepsilon^j~,
\\
\delta_{\hat\epsilon} \hat a_{\hat\mu} &= \frac{1}{2} i \xi^2 \left(\hat{\bar\psi}_{\hat\mu}^i\hat\varepsilon_i + \frac{2}{\sqrt{3}}\hat{\bar\chi}^i\hat\gamma_{\hat\mu}\hat\varepsilon_i \right)~,\\
\delta_{\hat\epsilon} \hat\chi_i&=\frac{1}{\sqrt{2}}\hat\gamma^{\hat\mu}\left(\partial_{\hat\mu}\hat\varphi\right)\hat\varepsilon_i+A_{ij}\hat\varepsilon^j
  -\frac{1}{2\sqrt{6}}\hat\gamma^{\hat\mu\hat\nu}
   \Hterm{\hat\mu\hat\nu ij}{}{-\sqrt{2}}\hat\varepsilon^j~,\\
\delta_{\hat\epsilon} \hat B_{\hat\mu\hat\nu}^\alpha&=
  2 \hat D_{[\hat\mu}\Theta_{\hat\nu]}^{ij}\left(\Gamma^\alpha\right)_{ij}
  -\frac{ig_1}{\sqrt{2}} \epsilon^{\alpha\beta}\left(\Gamma_\beta\right)_{ij}\xi
  \left(\hat{\bar\psi}_{[\hat\mu}^i\hat\gamma_{\hat\nu]}\hat\epsilon^j+\frac{1}{2\sqrt{3}}\hat{\bar\chi}^i\hat\gamma_{\hat\mu\hat\nu}\hat\epsilon^j\right)~,
\end{split}
\end{align}
where $\Theta_{\hat\mu}^{ij}=\sqrt{\frac{1}{2}}i\xi^{-1}\left(-\hat{\bar{\psi}}_{\hat\mu}^i\hat\epsilon^j+\sqrt{\frac{1}{3}}\hat{\bar{\chi}}^i\hat\gamma_{\hat\mu}\hat\epsilon^j\right)$.
The commutator of two supersymmetries is -- to leading order in the fermionic fields -- given by
\begin{align}\label{eqn:QQ-commutator}
 \left[\delta_{\hat\epsilon_2},\delta_{\hat\epsilon_1}\right] = \delta_{\hat X}+\delta_{\hat\Sigma}+\delta_{\hat\sigma}+\delta_{\hat\tau^I}~,
\end{align}
where $\delta_{\hat X}$ denotes a diffeomorphism with 
$\hat X^{\hat\mu}=-i\hat{\bar\epsilon}_1^i\hat\gamma^{\hat\mu}\hat\epsilon_{2 i}$\,,
$\delta_{\hat\Sigma}$ is a local Lorentz transformation with
\begin{align}
 \hat\Sigma^{\hat a\hat b} = \hat X^{\hat\mu}\hat\omega_{\hat\mu}^{\hphantom{\hat\mu}\hat a\hat b}
  +2i\hat{\bar\epsilon}^i_1\left(
   -\hat\gamma^{\hat a\hat b}T_{ij}+\frac{1}{6\sqrt{2}}\left(\hat\gamma^{\hat a\hat b}_{\ \ \,\hat c\hat d}+4\delta^{\hat a}_{\hat c}\delta^{\hat b}_{\hat d}\right)\Hterm{ij}{\hat c\hat d}{+\frac{1}{\sqrt{2}}}
  \right)\hat\epsilon_2^j~,
\end{align}
and $\delta_{\hat\sigma}$ and $\delta_{\hat\tau^I}$ denote U(1) and SU(2) gauge transformations, respectively, 
with 
\begin{align}
 \hat\sigma=\hat X^{\hat\mu}\hat a_{\hat\mu}+\frac{1}{2}i\xi^2\hat{\bar\epsilon}_1^i\hat\epsilon_{2i}~,
\qquad
\hat\tau^I=\hat X^{\hat \mu}\hat A_{\hat \mu}^I-\frac{1}{\sqrt{2}}i\xi^{-1}\left(\Gamma^I\right)_{ij}\hat{\bar\epsilon}_1^i\hat\epsilon_2^j~.
\end{align}

\section{Local \texorpdfstring{$\mathbf{\N{2}}$}{N=2} superconformal symmetry on the boundary
         of asymptotically-AdS configurations} \label{sec:boundary-limit}
We now restrict the configuration space of the theory discussed in the
previous section to geometries which are asymptotically AdS$_5$, 
and discuss the fields and symmetries induced on the conformal boundary.
We give a brief discussion of asymptotically-AdS spaces in the following, and refer to
\cite{Graham:1999jg, FeffermanGraham} for more details.
The metric signature and curvature conventions are those of Section~\ref{sec:RomansN4} and \cite{Romans:1985ps}, 
i.e.~AdS has positive curvature.

A metric $\hat g$ on the interior of a compact manifold $X$ with boundary $\partial X$
is called conformally
compact if, for a defining function $r$ of the boundary (meaning
that $r\vert_{\partial X}=0$, $dr\vert_{\partial X}\neq 0$ and $r\vert_{\text{int}X}> 0$), the rescaled
metric $\bar g:=r^2\hat g$ extends to all of $X$ as a metric.
For such a conformally compact metric $\hat g$ the  conformal structure 
$\left[\bar g\vert^{}_{T\partial X}\right]$ induced on $\partial X$ and the boundary restriction of the function 
$|dr|^2_{\bar g}:=\bar g^{\,-1}(dr,dr)$ 
are independent of the choice of defining function.
The curvature of the metric $\hat g$ is given by
\begin{align}\label{eqn:ghat-curvature}
 \hat{\calR}_{\hat\mu\hat\nu\hat\rho\hat\sigma}
 =-|dr|^2_{\bar g}\,\big(\hat g_{\hat\mu\hat\rho}\hat g_{\hat\nu\hat\sigma}
     -\hat g_{\hat\mu\hat\sigma}\hat g_{\hat\nu\hat\rho}\big)+\mathcal
     O(r^{-3})~,
\end{align}
where we denote tangent-space indices on $TX$ with hat, e.g.~$\hat\mu$,
$\hat\nu$, and
tangent-space indices on $T\partial X$ are denoted without hat.
Asymptotically, $\hat g$ thus has constant sectional curvature
given by $-|dr|^2_{\bar g}$, and we call a conformally compact metric 
$\hat g$ an asymptotically-AdS metric if the value of the sectional 
curvature is positive and constant on the boundary, 
i.e.~$|dr|^2_{\bar g}= -1/R^2$ on $\partial X$ for some constant $R$.
Note that we do not demand $\hat g$ to be Einstein.

A representative metric $g^{\sbr{0}}$ of the boundary conformal structure uniquely
determines a defining function $r$ such that $g^{\sbr{0}}=\frac{r^2}{R^2}\hat g\,\vert^{}_{T\partial X}$
and $|dr|^2_{\bar g}= -1/R^2$ in a neighbourhood of $\partial X$.
Choosing this defining function as radial coordinate, the metric $\hat g$ takes the
Fefferman-Graham form
\begin{align} \label{eqn:Fefferman-Graham}
  \hat g=\frac{R^2}{r^2}\left(g_{\mu\nu}dx^\mu\otimes dx^\nu
        -dr\otimes dr\right),
  \qquad
  g_{\mu\nu}(x,r)=g_{\mu\nu}^{\sbr{0}}(x)+\frac{r^2}{R^2}g_{\mu\nu}^{\sbr{2}}(x)+\dots
\end{align}
with $g$ of signature $(+,-,-,-)$ and
the limit $r\rightarrow 0$ corresponding to the conformal boundary.
The expansion of $g$ in powers of $r$ is justified when $\hat g$ satisfies vacuum
Einstein equations, which, however, we do not assume here.
For the time being we will still use that expansion and refer the discussion
of its validity to Section~\ref{sec:nonlinear-theory}.

According with the Fefferman-Graham form of the metric, we partially gauge-fix
the local Lorentz symmetry such that the vielbein is of the form
\begin{align}\label{eqn:vielbein-choice}
 \hat e_\mu^a(x,r)=\frac{R}{r} e_\mu^a(x,r),\qquad \hat
 e_\mu^{\underline{r}}=\hat e_r^a=0,\qquad \hat
 e_r^{\underline{r}}=\frac{R}{r}~,
\end{align}
with $e_\mu^a(x,r)=e_\mu^{\sbr{0}a}(x)+r e_\mu^{\sbr{1}a}(x)+\dots$~.
We denote Lorentz indices 
by $\hat a=(a,\underline{r})$ with an underline below $r$ to avoid confusion.
For the gravitinos and the
SU(2)$\otimes$U(1) gauge fields we employ axial gauges 
$\hat\psi_{ri}\equiv \hat A_r^I\equiv\hat a_r\equiv 0$.

In this setting we construct the fields induced on the conformal boundary
in Section~\ref{sec:boundary-fields}.
For the discussion of the induced symmetry transformations 
we will be interested in the residual
bulk symmetries preserving the gauge-fixing conditions.
These are to be determined as solutions to 
\begin{align}\label{eqn:gauge-fixing-constraints}
 \left(\delta_{\hat X}+\delta_{\hat\Sigma}+\delta_{\hat\epsilon_i}
       +\delta_{\text{U(1)}}+\delta_{\text{SU(2)}}\right)
  \lbrace \hat e_r^{\underline r}, \hat e_r^a,\hat e_\mu^{\underline{r}}, 
           \hat  a_r, \hat A_r^I, \hat \psi_{r i}\rbrace=0~,
\end{align}
where $\delta_{\hat X}$, $\delta_{\hat\Sigma}$, $\delta_{\hat \epsilon}$
denote diffeomorphisms, local Lorentz and supersymmetry transformations,
respectively.
The solutions and their action on the boundary fields will be discussed in
Section~\ref{sec:boundary-symmetries}.

The spin connection is treated in $1.5^\text{th}$-order formalism and 
fixed by its equation of motion as derived from (\ref{eqn:N4-Lagrangian1}).
We split 
$\hat\omega_{\hat\mu\hat a\hat b}=\hat\omega_{\hat\mu\hat a\hat b}(\hat e)+\hat\omega_{\hat\mu\hat a\hat b}(\hat e,\hat\psi,\hat\chi)$
where the torsion-free part 
$\hat\omega_{\hat\mu\hat a\hat b}\left(\hat e\right)$ 
calculated from (\ref{eqn:vielbein-choice}) 
has the non-vanishing components
\begin{align}\label{eqn:spin-connection-torsion-free}
\hat\omega_\mu^{\hphantom{\mu}ab}\left(\hat e\right)=\omega_\mu^{\hphantom{\mu}ab}(e)~,
\qquad 
\hat \omega_\mu^{\hphantom{\mu}a\underline{r}}\left(\hat e\right)=\frac{1}{r}e_\mu^a
     -\frac{1}{2}e^{\rho a}\partial_r g_{\mu\rho}
\qquad 
\hat \omega_r^{\hphantom{r}ab}\left(\hat e\right)=e^{\mu[a}\partial_r e_\mu^{b]}~,
\end{align}
and for the remaining part involving fermions we find
\begin{flalign}\label{eqn:spin-connection-torsion}
\hat\omega_{\hat\mu\hat a\hat b}(\hat e,\hat\psi,\hat\chi)
&=
  -\frac{1}{2}i\left(
    \hat{\bar\psi}_{\hat a}^i\hat\gamma_{\hat \mu}\hat\psi_{\hat b i}
    +2\hat{\bar\psi}_{\hat\mu}^i\hat\gamma_{[\hat a}\hat\psi_{\hat b] i}
   \right)
  -\frac{1}{4}i\hat{\bar\psi}_{\hat \lambda}^i\hat\gamma^{\ \ \ \ \hat\lambda\hat\tau}_{\hat\mu \hat a \hat b}\hat\psi_{\hat \tau i}
  -\frac{1}{4}i\hat{\bar\chi}^i\hat\gamma_{\hat \mu\hat a \hat b}\hat\chi_i~.
\end{flalign}
Thus, the Lorentz-covariant derivative on spinor fields reads
\begin{align}\label{eqn:covariant-derivative}
\begin{split}
\hat \nabla_\mu&=\nabla_\mu^\sbr{e}
       +\frac{1}{2r}\gamma_\mu\gamma_{\underline{r}}
       -Z_\mu
       +\frac{1}{4}\hat\omega_{\mu}^{\hphantom{\mu}\hat a\hat b}(\hat e,\hat\psi,\hat\chi)\hat\gamma_{\hat a\hat b}
        \ \ =:\nabla_\mu+\frac{1}{2r}\gamma_\mu\gamma_{\underline{r}}~,\\
\hat \nabla_r&=\partial_r
              -Z_r
              +\frac{1}{4}\hat\omega_{r}^{\hphantom{r}\hat a\hat b}(\hat e,\hat\psi,\hat\chi)\hat\gamma_{\hat a\hat b}~,
\end{split}
\end{align}
where $\hat\gamma_{\hat \mu}=\hat e_{\hat\mu}^{\hat a}\gamma_{\hat a}$\,, $\gamma_\mu=e_\mu^a\gamma_a$.
For notational convenience we defined
$\nabla^\sbr{e}_\mu:=\partial_\mu+\frac{1}{4}\omega_{\mu}^{\hphantom{\mu}ab}(e)\gamma_{ab}$
and
$Z_\mu:=\frac{1}{4}\left(\partial_r g_{\mu\rho}\right) \gamma^\rho\gamma_{\underline{r}}$~,
$Z_r:=\frac{1}{4}(\partial_r e^a_\mu)\gamma_{a}^{\ \,\mu}$\,.

\subsection{Boundary fields} \label{sec:boundary-fields}
In this section we construct the fields induced on the conformal boundary.
Similar to the construction
of the induced conformal structure on the boundary,
we define the 
classical boundary field as follows.
For a bulk field $\hat\phi$ with asymptotic $r$-dependence 
$\hat\phi(x,r)=\mathcal O(f(r))$, 
we define the rescaled field $\phi(x,r):=f(r)^{-1}\hat\phi(x,r)$.
This rescaled field then admits a finite, nonvanishing boundary limit, 
which is interpreted as the boundary field%
\footnote{This is the classical analog to the construction for the Wightman field in \cite{Bertola:2000pp, Rehren:2004yu}.
}.

Therefore, to determine the multiplet of boundary fields, we have to fix the asymptotic scaling
of the various fields. 
To this end we consider their equations of motion linearized in
all fields but the metric/vielbein and decomposed into 
boundary-irreducible components, 
e.g.~into four-dimensional chiral components for a bulk spinor field.
The leading order in the boundary limit turns out to be an ordinary differential equation
in $r$, and is solved by fixing the scalings of the different boundary-irreducible bulk field components.
The rescaled field is defined by extracting the asymptotic $r$-dependence of the dominant field component, 
thereby subdominant components are projected out in the definition of the boundary field.
The results obtained in this way on the basis of the linearized field equations are extended to the 
nonlinear theory in Section~\ref{sec:nonlinear-theory}.

We start with the vielbein, for which the asymptotic $r$-dependence is already fixed by 
(\ref{eqn:Fefferman-Graham}), (\ref{eqn:vielbein-choice}) 
and the induced boundary field is given by $e_\mu^a(x,0)$. 
As discussed in \cite{Romans:1985ps}, Einstein's equations as derived from (\ref{eqn:N4-Lagrangian1})
in a pure metric-dilaton background read
\begin{align}\label{eqn:Einstein-N4}
 \hat\calR_{\hat\mu\hat\nu}-\frac{1}{2}\hat g_{\hat\mu\hat\nu}\hat\calR+2\hat g_{\hat\mu\hat\nu}P(\hat\varphi)=0~,
\end{align}
and the scalar potential $P(\hat\varphi)$, having exactly one extremum 
$\big(\hat\varphi,P(\hat\varphi)\big)\equiv\left(0,\frac{3}{8}g^2\right)$,
provides a cosmological constant such that AdS$_5$ is a vacuum solution.
Here we do not restrict the theory to the metric-dilaton sector 
and only demand (\ref{eqn:Einstein-N4}) to be solved at leading order in the boundary limit.
From (\ref{eqn:ghat-curvature}) we find that $\hat g$ indeed solves the leading order provided that the
asymptotic curvature radius $R$ is fixed in terms of the gauge coupling as $R^2=8/g^2$.
In Section~\ref{sec:nonlinear-theory} we show that -- with the scalings obtained in this
section -- all other terms in the complete Einstein equations contribute to the subleading orders only.
In the following we fix $g=2\sqrt{2}$ such that $R=1$.

For the gravitinos, which we consider next, 
the nonlinear equation of motion reads
\begin{align}\label{eqn:gravitino-nonlinear}
\begin{split}
 \hat\gamma^{\hat\mu\hat\nu\hat\rho}\hat D_{\hat\nu}\hat\psi_{\hat\rho i}
 -3 T_{ij}\hat\gamma^{\hat\mu\hat\nu}\hat\psi_{\hat\nu}^j
 =&
 -\frac{1}{2\sqrt{2}}\Hterm{\ \ \,i}{\hat\rho\hat\sigma\ \,j}{+\frac{1}{\sqrt{2}}}
   \hat\gamma^{[\hat\mu}\hat\gamma_{\hat\rho\hat\sigma}\hat\gamma^{\hat\nu]}\hat\psi_{\hat\nu j}
 -A_{ij}\hat\gamma^{\hat\mu}\chi^j
 \\
 &-\frac{1}{2\sqrt{6}}\Hterm{\hat\rho\hat\sigma i}{\ \ \ \ j}{-\sqrt{2}}\hat\gamma^{\hat\rho\hat\sigma}\hat\gamma^{\hat\mu}\chi_j 
  +\frac{1}{\sqrt{2}}\left(\partial_{\hat\nu}\hat\varphi\right)\hat\gamma^{\hat\nu}\hat\gamma^{\hat\mu}\chi_i~.
\end{split}
\end{align}
To fix $T_{ij}$ (see (\ref{eqn:defTij})) we note that, since it squares to $-\mathds{1}$ and is traceless,  
$\Gamma_{45}$ has eigenvalues $\pm i$, each with multiplicity 2.
We choose a $\mathfrak{usp}(4)$ basis where $\Gamma_{45}$ is diagonal 
$\left(\Gamma_{45}\right)_i^{\ j}=i \kappa_i\delta_{i}^{\ j}$ 
and split $i=(i_+,i_-)$ such that $\kappa_{i_\pm}=\pm1$.
Since $\Gamma_{45}$ is diagonal  $\lbrace\Omega,\Gamma_{45}\rbrace=0$,
and consequently $\Omega^{i_+j_+}=\Omega^{i_-j_-}=0$.
Defining four-dimensional chirality projectors
$P_\text{L/R}:=\frac{1}{2}\left(1\pm i\gamma^{\underline r}\right)$, 
the L/R projections
of the linearized equation (\ref{eqn:gravitino-nonlinear}) for $\hat\mu=\mu$ read
\begin{align}\label{eqn:gravitino-lin}
\gamma^{\mu\nu\rho}\nabla^\sbr{e}_\nu\hat\psi_{\rho i}^\text{R/L}
 -\left( \gamma^{\mu\nu\rho}Z_\nu
 \pm i\gamma^{\mu\rho} Z_r \right)\hat\psi_{\rho i}^\text{L/R}
 +i\gamma^{\mu\rho}
  \left(\pm\partial_r \mp \frac{1}{r}+\frac{3 \kappa_i}{2r}\right)\hat\psi_{\rho i}^\text{L/R}=0~.
\end{align}
Since the $\hat\psi_{\mu i_-}^\text{L/R}$ are related to the conjugates of
$\hat\psi_{\mu i_+}^\text{R/L}$ by the 
symplectic Majorana condition, it is sufficient to consider the $i_+$-components.
Solving (\ref{eqn:gravitino-lin}) at leading order in $r$ yields the two independent
solutions 
$\hat\psi_{\mu i_+}=r^{-1/2}\psi_{\mu i_+}^\text{L}+\calo(r^{-1/2})$
and
$\hat\psi_{\mu i_+}=r^{5/2}\psi_{\mu i_+}^\text{R}+\calo(r^{5/2})$
with $\lim_{r\rightarrow 0}\psi_{\mu i_+}^\text{L/R}$ finite.
Thus, the gravitinos lose half of their components in the boundary limit
and the rescaled field $\psi_{\mu i_+}:=r^{1/2}\hat\psi_{\mu i_+}$
yields the two chiral gravitinos
$\psi_{\mu i_+}^\text{L}\vert^{}_{r=0}$ 
as boundary fields.

Proceeding with the fermionic fields we now discuss the spin-$\frac{1}{2}$ fermions $\hat\chi_i$. 
Their equation of motion is given by
\begin{align}\label{eqn:spin-half-eom}
\begin{split}
 \hat\gamma^{\hat\mu}\hat D_{\hat \mu}\hat\chi_i + T_{ij}\hat\chi^j
=\:&
\frac{2}{\sqrt{3}}A_{ij}\hat\chi^j
+A_{ij}\hat\gamma^{\hat\mu}\hat\psi_{\hat\mu}^j
+\frac{1}{2\sqrt{6}} \Hterm{\hat\mu\hat\nu i}{\ \ \ \ j}{-\sqrt{2}} \hat\gamma^{\hat\rho}\hat\gamma^{\hat\mu\hat\nu}\hat\psi_{\hat\rho j}
\\
&-\frac{1}{6\sqrt{2}} \Hterm{\hat\mu\hat\nu i}{\ \ \ \ j}{-\frac{5}{\sqrt{2}}} \hat\gamma^{\hat\mu\hat\nu}\hat\chi_j
+\frac{1}{\sqrt{2}}\left(\partial_{\hat\nu}\hat\varphi\right)\hat\gamma^{\hat\mu}\hat\gamma^{\hat\nu}\hat\psi_{\hat\mu i}~.
\end{split}
\end{align}
Solving the linearized L/R projections given by
 \begin{align}
  \gamma^\mu \nabla^\sbr{e}_\mu\hat\chi_i^\text{R/L}-\left(\gamma^\mu Z_\mu\mp i Z_r\right)\hat\chi_i^\text{L/R}
  -i\left(\pm\partial_r+\frac{\kappa_i\mp 4}{2r}\right)\hat\chi_i^\text{L/R}=0
 \end{align}
at leading order for $i=i_+$
we find as dominant solution
$\hat\chi^{}_{i_+}=r^{3/2}\chi_{i_+}^\text{L}+\calo(r^{3/2})$.
Similarly to the gravitinos, the $\hat\chi^{}_{i_+}$ become chiral in the boundary limit and we have
the two lefthanded Weyl fermions $\chi_{i_+}^\text{L}\vert^{}_{r=0}$ as boundary fields.

Coming to the tensor fields $\hat B_{\hat\mu\hat\nu}^\alpha$ we define
$\hat C_{\hat\mu\hat\nu}:=\frac{1}{\sqrt{2}}(\hat B^4_{\hat\mu\hat\nu}-i\hat B^5_{\hat\mu\hat\nu})$
and, 
with the four-dimensional Hodge dual 
$\star\,\hat C_{\mu\nu}:=\frac{1}{2}e^{-1}\epsilon_{\mu\nu}^{\ \ \ \rho\sigma}\hat C_{\rho\sigma}$, 
the (anti-)selfdual parts of $\hat C_{\mu\nu}$ are defined as
$\hat C_{\mu\nu}^\pm:=\frac{1}{2}(\hat C_{\mu\nu}\pm i \star \hat C_{\mu\nu})$.
The equation of motion reads
\begin{align}\label{eqn:Ctensor-eom}
\frac{i}{g_1}\hat\epsilon^{\hat\mu\hat\nu\hat\rho\hat\sigma\hat\tau}\hat D_{\hat\rho}\hat C_{\hat\sigma\hat\tau}
-\hat e\xi^2\hat C^{\hat\mu\hat\nu}
=
-\frac{1}{2} \hat e \xi\Big( &
 \frac{1}{2} {J_1}_{ij}^{\hat\mu\hat\nu}
 + \frac{1}{\sqrt{3}}{J_2}_{ij}^{\hat\mu\hat\nu}
 - \frac{1}{6}{J_3}_{ij}^{\hat\mu\hat\nu}
 \Big)\left(\Gamma_4-i\Gamma_5\right)^{ij}~,
\end{align}
with 
${J_1}_{ij}^{\hat\mu\hat\nu}=i\hat{\bar{\psi}}_i^{\hat\rho}\hat\gamma_{[\hat\rho}\hat\gamma^{\hat\mu\hat\nu}\hat\gamma_{\hat\sigma]}\hat{\psi}_j^{\hat\sigma}$\,,\,
${J_2}_{ij}^{\hat\mu\hat\nu}=i\hat{\bar\psi}_i^{\hat\rho} \hat\gamma^{\hat\mu\hat\nu}\hat\gamma_{\hat\rho}\hat\chi_j$
and
${J_3}_{ij}^{\hat\mu\hat\nu}=i\hat{\bar\chi}_i\hat\gamma^{\hat\mu\hat\nu}\hat\chi_j$.
From the $\mu r$-compo\-nents of the linearized equation 
$\hat C_{\mu r}$
is fixed in terms of $\hat C_{\mu\nu}$ by
$\hat C_{\mu r}=\frac{1}{2}i r e^{-1}\epsilon_{\mu}^{\mbox{\ \,}\rho\sigma\tau}\partial_\rho\hat C_{\hat\sigma\hat\tau}$, 
and is of higher order in $r$.
The (anti-)selfdual parts of the linearized $\mu\nu$-components
\begin{align}\label{eqn:Cmunu-lin-eom}
\frac{1}{2}e^{-1}\epsilon_{\mu\nu}^{\ \ \ \rho\sigma}\left(\partial_r \hat C_{\mu\nu}+2\partial_\rho\hat C_{\sigma r}\right)=-\frac{i}{r} \hat C_{\mu\nu}~,
\end{align}
then yield the solutions
$\hat C_{\mu\nu}=r^{-1}C^-_{\mu\nu}+\calo(r^{-1})$ 
and 
$\hat C_{\mu\nu}=r\,C^+_{\mu\nu}+\calo(r)$.
Thus, the anti-selfdual part $\hat C^-$ is dominant in the boundary limit 
and the selfdual part $\hat C^+$ is projected out in the definition of the boundary field.

For the U(1) and SU(2) gauge fields the equations of motion are
\begin{align}
\begin{split}\label{eqn:eom-amu}
\partial_{\hat\nu}\left(\hat e\xi^{-4}\hat f^{\hat\mu\hat\nu}\right)
=\:&
\frac{1}{4}\hat e g_1\left(\Gamma_{45}\right)_{i}^{\ j}
{J_4}^{\hat\mu i}_{\ \ j}
-\frac{1}{4}\hat\epsilon^{\hat\mu\hat\nu\hat\rho\hat\sigma\hat\tau}
  \left(\hat{B}^\alpha_{\hat\nu\hat\rho}\hat B^\alpha_{\hat\sigma\hat\tau}
         +\hat F_{\hat\nu\hat\rho}^I \hat F_{\hat\sigma\hat\tau}^I\right)
\\
&+\Omega^{ij}\partial_{\hat\nu}\left(\hat e \xi^{-2}\left(
\frac{1}{4}{J_1}^{\hat\mu\hat\nu}_{ij}-\frac{1}{\sqrt{3}}{J_2}^{\hat\mu\hat\nu}_{ij}+\frac{5}{12}{J_3}^{\hat\mu\hat\nu}_{ij}
\right)\right)~,
\end{split}
\\
\label{eqn:eom-AImu}
\hat D_{\hat\nu}\left(\hat e\xi^{2}\hat F^{I \hat\mu\hat\nu}\right)
=\:&
\frac{1}{4}\hat e g_2 \left(\Gamma_{I45}\right)_{i}^{\ j} {J_4}^{\hat\mu i}_{\ \ j}
-\hat \epsilon^{\hat\mu\hat\nu\hat\rho\hat\sigma\hat\tau} \hat D_{\hat\nu}\left(\hat F_{\hat\rho\hat\sigma}^I\hat a_{\hat\tau}\right)
+\frac{1}{\sqrt{2}} \hat D_{\hat\nu}\left(\hat e \xi K_{I}^{\hat\mu\hat\nu}\right)~,
\end{align}
with ${J_4}^{\hat\mu i}_{\ \ j}=i\hat{\bar\chi}^i\hat\gamma^{\hat\mu}\hat\chi_j-i\hat{\bar{\psi}}_{\hat\nu}^i\hat\gamma^{\hat\nu\hat\mu\hat\rho}\hat\psi_{\hat\rho j}$
and
$K_{I}^{\hat\mu\hat\nu}=\left(\Gamma_I\right)^{ij}
\left(\frac{1}{2}{J_1}_{ij}^{\hat\mu\hat\nu}+\frac{1}{\sqrt{3}}{J_2}_{ij}^{\hat\mu\hat\nu}-\frac{1}{6}{J_3}_{ij}^{\hat\mu\hat\nu}\right)$.
For the ansatz $\hat a_\mu=r^{\alpha}a_\mu$
the leading order of the linearized equation yields $\alpha\in\lbrace 0,2\rbrace$,
and similarly for $\hat A_\mu^I$.
Thus, $\hat a_{\mu}$ and $\hat A_{\mu}$ are itself finite in the boundary limit and
define boundary vector fields without rescaling. 

It remains to analyze the scalar field $\hat\varphi$ with equation of motion
\begin{align}\label{eqn:scalar-nonlinear}
\begin{split}
\hat{\square}_{\hat g}\hat\varphi-P^\prime(\hat\varphi)
=& 
-\frac{i}{\sqrt{2}} A_{ij}\hat{\bar\psi}_{\hat\mu}^i\hat\gamma^{\hat\mu\hat\nu}\hat\psi_{\hat\nu}^j
-iA^\prime_{ij}\hat{\bar\psi}_{\hat\mu}^i\hat\gamma^{\hat\mu}\hat\chi^j
-\frac{i}{\sqrt{3}}\Big(A^\prime_{ij}+\frac{1}{\sqrt{6}}A_{ij}\Big)\hat{\bar\chi}^i\hat\chi^j
\\ &
-\sqrt{\frac{2}{3}}\xi^2\hat{\overline{C}}_{\hat\mu\hat\nu} \hat C^{\hat\mu\hat\nu}
+\sqrt{\frac{2}{3}}\xi^{-4}\hat f_{\hat\mu\hat\nu}\hat f^{\hat\mu\hat\nu}
-\frac{1}{\sqrt{6}}\xi^2\hat F^I_{\hat\mu\hat\nu}\hat F^{I\hat\mu\hat\nu}
\\ &
+\frac{1}{4\sqrt{3}}\Hterm{\hat\mu\hat\nu}{ij}{-\sqrt{2}}{J_1}_{ij}^{\hat\mu\hat\nu}
+\frac{1}{6}\Hterm{\hat\mu\hat\nu}{ij}{+2\sqrt{2}}{J_2}_{ij}^{\hat\mu\hat\nu}
\\ &
-\frac{1}{12\sqrt{3}}\Hterm{\hat\mu\hat\nu}{ij}{+5\sqrt{2}}{J_3}_{ij}^{\hat\mu\hat\nu}
-\frac{1}{\sqrt{2}}\hat e^{-1}\partial_{\hat\nu}\left(i\hat e \hat{\bar\psi}_{\hat\mu}^i\hat\gamma^{\hat\nu}\hat\gamma^{\hat\mu}\hat\chi_i\right)~,
\end{split}
\end{align}
where ${A^\prime}^{ij}:=-\frac{1}{6}g\left(\xi^{-1}+2\xi^2\right)\left(\Gamma_{45}\right)^{ij}$.
The linearized equation is given by
\begin{align}
 r^2\square_g\hat\varphi-\frac{1}{2}r^2\left(g^{\mu\nu}\partial_r g_{\mu\nu}\right)\partial_r \hat\varphi
 -\mathcal D_r^2\hat\varphi=0~,
\end{align}
with $\mathcal D_r=r\partial_r-2$, and 
the leading-order part is solved by  
$r^2\varphi_1(x,r)$ and  $r^2\log(r)\,\varphi_2(x,r)$ with $\varphi_{1/2}\vert^{}_{r=0}$ finite.
The boundary scalar field is thus defined by extracting the dominant scaling 
$\hat\varphi=:r^2\log(r)\varphi$ and restricting $\varphi$ to the boundary.
In summary, the multiplet of boundary fields is given 
by $(e_\mu^a, \psi_{\mu\,i_+}^\text{L},C_{\mu\nu}^-,A_\mu^I,a_\mu,\chi_{i_+}^\text{L},\varphi)\vert^{}_{r=0}$.

\subsection{Nonlinear theory and subdominant components} \label{sec:nonlinear-theory}
The splitting into dominant and subdominant components and the 
scaling of the dominant parts as obtained above from the linearized equations of motion 
fixes the definition of the boundary fields.
It remains to be checked whether the obtained scaling behaviour is consistent in the nonlinear theory.
Furthermore, the subdominant components of some of the fields are required for the symmetry transformations 
to be discussed in Section~\ref{sec:boundary-symmetries}.
These two points are addressed in the following.
Note that this discussion does not include the four-fermion terms which are not spelled out in \cite{Romans:1985ps}.
However, as we find quite some cancellations taking place to ensure 
consistency of the previously obtained results at the leading orders in the fermions, 
we expect that this consistency is not accidental and extends to the four-fermion terms as well.

Since the analysis of Section~\ref{sec:boundary-fields} crucially relies on the
form of the metric (\ref{eqn:Fefferman-Graham}) in a neighbourhood of the boundary, 
the first thing to be checked is the validity of the Fefferman-Graham form.
Considering the terms in the Lagrangian (\ref{eqn:N4-Lagrangian1}) with the scaling of the 
fields as obtained in the previous section, 
$\hat e \hat {\calR}(\hat\omega)$ and the cosmological constant $\hat e P(0)$ are $\mathcal O(r^{-5})$ 
while the other terms are $\mathcal O(r^{-3})$. 
Thus, the leading order of Einstein's equations reduces to the form discussed 
in the previous section and the Fefferman-Graham form of the metric (\ref{eqn:Fefferman-Graham}) is justified.
In particular, since there are no $\mathcal O(r^{-4})$ terms in the Lagrangian,
there is no $\mathcal O(r)$ contribution to $g_{\mu\nu}(x,r)$ and the expansion in (\ref{eqn:Fefferman-Graham}) is justified.
Next, we consider the spin connection 
(\ref{eqn:spin-connection-torsion-free}), (\ref{eqn:spin-connection-torsion}).
With the scaling as obtained before, $\hat\omega_{\mu a b}(\hat e,\hat\psi,\hat\chi)=\mathcal O(r^0)$ and
the other components of $\hat\omega_{\hat\mu \hat a \hat b}(\hat e,\hat\psi,\hat\chi)$ 
are of $\mathcal O(r)$.
Therefore, the fermionic terms do not alter the $\mathcal O(r^{-1})$ part of the covariant derivative
(\ref{eqn:covariant-derivative}), which was relevant for the previous section.
For the four-dimensional Lorentz-covariant 
derivative $\nabla_\mu$ defined in (\ref{eqn:covariant-derivative}) we find 
$\nabla_\mu=\partial_\mu+\frac{1}{4}\omega_{\mu}^{\hphantom{\mu}ab}\gamma_{ab}$ with 
\begin{align}\label{eqn:spin-connection-4d}
 \omega_{\mu a b}\big\vert_{r=0}&=\omega_{\mu a b}(e)
       -\frac{1}{2}\left(
    i{\bar\psi}^{\text{L} i_+}_{a}\gamma_{\mu}\psi_{b i_+}^\text{L}
    +2i{\bar\psi}_{\mu}^{\text{L} i_+}\gamma_{[ a}\psi_{ b] i_+}^\text{L}
   +\text{c.c.}\right)~.
\end{align}
From (\ref{eqn:covariant-derivative-spinor4}) the four-dimensional gauge and Lorentz covariant 
derivative acting on a boundary spinor is
\begin{align}\label{eqn:covariant-derivative-4d}
D_{\mu} v_{i_+}=\nabla_{\mu} v_{i_+} + \frac{1}{2}i g_1  a_{\mu} v_{i_+} + \frac{1}{2} i g_2 A_{\mu}^I\left(\Gamma_{I}\right)_{i_+}^{\ \ {j_+}}v_{j_+}~.
\end{align}

For the remaining fields we study the interaction terms of 
(\ref{eqn:N4-Lagrangian1}) directly in the field equations.
They turn out to be subdominant in the equations for the boundary-dominant 
field components, such that their scaling is not affected.
They do, however, alter the subdominant components, some of which are in fact 
not subdominant but play the role of auxiliary fields on the boundary.
We start with the gravitinos, for which the 
scaling of $\hat\psi_{\mu i_+}^\text{L}$ was determined from the $P_\text{L}$-projection of (\ref{eqn:gravitino-nonlinear}) at $\mathcal O(r^{3/2})$. 
One easily verifies that the interaction terms in (\ref{eqn:gravitino-nonlinear}) are of $\mathcal O(r^{5/2})$ and thus 
the analysis of the previous section is not affected. 
To determine the subdominant components we consider the $P_\text{R}$ 
projection of the $\hat\mu{=}\mu$ components. 
Noting that
$\left(\Gamma_\alpha\right)_{i_+}^{\ \,j_+}=\left(\Gamma_\alpha\right)_{i_-}^{\ \,j_-}=0$ 
due to
$\lbrace\Gamma_\alpha,\Gamma_{45}\rbrace=0$,
and 
$\hat B_{\hat\mu\hat\nu}^\alpha\left(\Gamma_\alpha\right)_{i_+}^{\ \,j_-}=\sqrt{2}\hat C_{\hat\mu\hat\nu}\left(\Gamma_4\right)_{i_+}^{\ \,j_-}$,
we find
$\hat\psi_{\mu i_+}=r^{-1/2}\psi_{\mu i_+}^\text{L}+r^{1/2}\Phi_{\mu i_+}^\text{R}$ with
\begin{align} \label{eqn:gravitino-subdominant}
 \Phi_{\mu i_+}^\text{R} \Big\vert_{r=0}&=   
     -\frac{1}{2}i\Big(\gamma_{\mu}^{\ \,\nu\rho}-\frac{2}{3}\gamma_{\mu}\gamma^{\nu\rho}\Big)
      \Big(D_\nu\psi_{\rho i_+}^\text{L} -\frac{1}{4}\gamma\cdot C_{i_+j_+}^-\gamma_{\nu}\psi_{\rho}^{\text{R}j_+}\Big)
      ~,
\end{align}
where $\gamma\cdot C:=\gamma^{\mu\nu}C_{\mu\nu}$ and $C_{\hat\mu\hat\nu\,i_+ j_+}:=C_{\hat\mu\hat\nu}\left(\Gamma_4\right)_{i_+j_+}$.
Note that $\psi_\mu^{\text{R}i_+}=C\big(\bar\psi_{\mu}^{\text{L}i_+}\big)^T$ by the symplectic Majorana condition, and 
a possible $C^+$-contribution drops out 
due to $\gamma\cdot C^\pm = \gamma\cdot C^\pm P_\text{R/L}$. 
For later convenience we define the quantity
\begin{align}
 R_{\mu\nu\,i_+}(Q) := D_{[\mu}\psi_{\nu]i_+}^\text{L}-i \gamma_{[\mu}\Phi_{\nu]i_+}^\text{R}-\frac{1}{4}\gamma\cdot C^-_{i_+ j_+}\gamma_{[\mu}\psi_{\nu]}^{\text{R}j_+}~,
\end{align}
and note that it is anti-selfdual $i\star R_{\mu\nu\,i_+}(Q)=-R_{\mu\nu\,i_+}(Q)$ and satisfies $\gamma^\mu R_{\mu\nu\,i_+}(Q)=0$.

We continue with the tensor field $\hat C_{\hat\mu\hat\nu}$.
Using $\frac{1}{2}\left(\gamma_{\mu\nu}\pm i \star \gamma_{\mu\nu}\right)=\gamma_{\mu\nu}P_\text{R/L}$ we 
find the interaction terms subdominant in the anti-selfdual part of (\ref{eqn:Ctensor-eom}) with $\hat\mu\hat\nu{=}\mu\nu$,
which was used to determine the scaling  $\hat C^-_{\mu\nu}=r^{-1}C_{\mu\nu}^-$.
In the selfdual part the interaction terms are not subdominant,
but rather fix $\hat C_{\mu\nu}^+= r^{-1}C_{\mu\nu}^+$ with
\begin{align}\label{eqn:Cplus}
 C_{\mu\nu}^+\big\vert_{r=0}&=
    \frac{1}{4}i\left(\Gamma_4\right)^{i_+ j_+}\bar\psi_{\rho i_+}^{\text{R}}\gamma^{[\rho}\gamma_{\mu\nu}\gamma^{\sigma]}\psi_{\sigma j_+}^\text{L}~.
\end{align}
Thus, $\hat C_{\mu\nu}^+$ is in fact not subdominant with respect to $\hat C_{\mu\nu}^-$. 
However, since its boundary value  is completely fixed in terms of the other boundary fields, $\hat C_{\mu\nu}^+$ 
plays the role of an auxiliary field on the boundary.
From the $\hat\mu \hat\nu{=}\mu r$ components we find the subdominant $\hat C_{\mu r}=C_{\mu r}$ with
\begin{align} \label{eqn:Cmur}
 C_{\mu r} \big\vert_{r=0} &= 
     \frac{1}{2}ir e^{-1}\epsilon_{\mu}^{\ \ \nu\rho\sigma}D_\nu \hat C_{\rho\sigma}
      +\bar\psi_{\rho i_+}^\text{R}\big(\gamma_\mu^{\ \,\rho\sigma}\Phi_{\sigma j_+}^\text{R}+\frac{1}{\sqrt{3}}\gamma_\mu\gamma^\rho\chi_{j_+}^\text{L}\big)\left(\Gamma_4\right)^{i_+j_+}~.
\end{align}

For the spin-$\frac{1}{2}$ fermions $\hat\chi_{i_+}^\text{L}=r^{3/2}\chi_{i_+}^\text{L}$ was obtained from
the $P_\text{L}$ projection of (\ref{eqn:spin-half-eom}) at $\mathcal O(r^{3/2})$.
The only additional contribution at that order is $\propto \gamma^\rho\gamma\cdot C_{i_+j_+}^+\psi_{\rho}^{\text{R}j_+}$
which is a three-fermion term by (\ref{eqn:Cplus}) and we expect it to be
cancelled by contributions of four-fermion terms in (\ref{eqn:N4-Lagrangian1}).
We conclude that -- up to the four-fermion terms not considered here -- the obtained scaling 
for $\hat\chi_{i_+}^\text{L}$ is not affected by the interaction terms.
The subdominant righthanded part is fixed from the
$P_\text{R}$-projection of (\ref{eqn:spin-half-eom})
and we find 
$\hat\chi^{}_{i_+}=r^{3/2}\chi_{i_+}^\text{L}+r^{5/2}\log(r)\chi_{i_+}^\text{R}$
with
\begin{align} \label{eqn:chi-subdominant}
\begin{split}
 \chi_{i_+}^\text{R}\Big\vert_{r=0}=i\slashed{D}\chi_{i_+}^\text{L}
  &-\frac{1}{\sqrt{2}}\varphi\gamma^\mu\psi_{\mu i_+}^\text{L}
   -\frac{i}{2\sqrt{6}}\gamma^\rho\gamma^{\mu\nu} 
   \left(F_{\mu\nu}^I\left(\Gamma_I\right)_{i_+}^{\ \,j_+}-\sqrt{2}f_{\mu\nu}\delta_{i_+}^{\ \,j_+}\right)\psi_{\rho j_+}^\text{L}
  \\ &
  +\frac{i}{2\sqrt{3}}\gamma^\rho\gamma\cdot C_{i_+ j_+}^-\Phi_\rho^{j_+ \text{L}}
  -\frac{1}{\sqrt{3}}\gamma^\rho\gamma^\mu C_{\mu r\, i_+ j_+}\psi_\rho^{j_+\text{R}}~.
\end{split}
\end{align}

In the equations for the gauge fields
(\ref{eqn:eom-amu}), (\ref{eqn:eom-AImu})
the leading-order terms are those involving 
${J_4}^{\hat\mu i}_{\ \ j}$ (the gravitino part thereof)
and ${J_1}^{\mu\nu}_{ij}$, both of which are of $\mathcal O(r^{-3})$.
However, since 
$\left(\Gamma_I\right)_{i_+}^{\ \,j_-}=\left(\Gamma_I\right)_{i_-}^{\ \,j_+}=0$ 
due to $\left[\Gamma_I,\Gamma_{45}\right]=0$,
their leading-order parts cancel exactly in both equations, 
such that the previous analysis of the linearized equations is not altered.
For the scalar field we have to check that the interaction terms are subdominant with respect to the
$\mathcal O(r^2)$ and $\mathcal O(r^2\log(r))$ parts of (\ref{eqn:scalar-nonlinear}).
Similar to the case of the gauge fields, there are cancellations between different terms at leading order.
From (\ref{eqn:Cplus}) the $J_{1ij}^{\mu\nu}$ term and the $\hat{\overline{C}}_{\mu\nu}\hat{C}^{\mu\nu}$ term add up 
to zero at leading order,
and also $-iA^\prime_{ij}\hat{\bar\psi}_{\hat\mu}^i\hat\gamma^{\hat\mu}\hat\chi^j$ 
and 
$-\frac{1}{\sqrt{2}}\hat e^{-1}\partial_{\hat\nu}\left(i\hat e \hat{\bar\psi}_{\hat\mu}^i\hat\gamma^{\hat\nu}\hat\gamma^{\hat\mu}\hat\chi_i\right)$
cancel.
The remaining terms are subleading and thus the cancellations justify 
the analysis of the linearized equations also for $\hat\varphi$.
We conclude that the scaling behaviours obtained from the linearized equations of motion 
with the modifications for the subdominant components discussed here
are consistent in 
the nonlinear theory as given by (\ref{eqn:N4-Lagrangian1}).

\subsection{Induced boundary symmetries}\label{sec:boundary-symmetries} 
Having obtained the multiplet of boundary fields in the previous section we now discuss the
symmetries on the boundary.
To this end we determine the residual bulk symmetries from the constraints 
(\ref{eqn:gauge-fixing-constraints})
and examine their action on the boundary fields, which is defined straightforwardly 
e.g.~$\delta\phi:=\lim_{r\rightarrow 0}f(r)^{-1}\hat\delta\hat\phi$
for a boundary field $\phi=\lim_{r\rightarrow 0}f(r)^{-1}\hat\phi$.
Relevant to us are solutions to the constraints (\ref{eqn:gauge-fixing-constraints}) 
which act nontrivially on the boundary fields,
and in the following we discuss certain special solutions which generate the general symmetry transformation of the
boundary fields.

The constraint that $\hat e_r^{\underline r}$ and $\hat e_r^a$ 
be preserved yields that, for an arbitrary $\lambda(x)$,
\begin{align}\label{eqn:constraints1}
\hat X^{r}=r\lambda(x),\qquad  \hat\Sigma^{a}_{\ \,\underline r}=-e_\mu^a\partial_r \hat X^\mu~.
\end{align}
We parametrize the U(1) and SU(2) gauge transformations by $\hat\sigma(x,r)$ and $\hat\tau^I(x,r)$, respectively, 
and 
using (\ref{eqn:constraints1}) the remaining constraints are
\begin{subequations}\label{eqn:constraints2}
\begin{align}
 \partial_r \hat X^\mu =&g^{\mu\rho}\big(r\partial_\rho \lambda(x)+i \hat{\bar \psi}_\rho^i\hat\gamma^{r}\hat\epsilon_i\big)~,
 \label{eqn:delta-e-constraint}
 \\
  \partial_r \hat\sigma=&\hat a_\mu \partial_r \hat X^\mu+\frac{1}{\sqrt{3}}i\xi^2\hat{\bar\chi}^i\hat\gamma_r\hat\epsilon_i~,
 \label{eqn:deltaU1constraint}
 \\
 \partial_r\hat\tau^I=&\hat A_\mu^I\partial_r \hat X^\mu+\frac{1}{\sqrt{6}}i\xi^{-1}\hat{\bar\chi}^i\hat\gamma_r\hat\epsilon^j\left(\Gamma^I\right)_{ij}~,
 \label{eqn:deltaSU2constraint}
\\
\begin{split}
 \hat\nabla_r\hat\epsilon_i+\hat\gamma_r T_{ij}\hat\epsilon^j
 =&
 - \big(\partial_r \hat X^\mu\big)\hat\psi_{\mu i}
 +\frac{1}{6\sqrt{2}}\Big(\hat\gamma_{r}^{\ \hat\nu\hat\rho}-4\delta_{r}^{\hat\nu}\hat\gamma^{\hat\rho}\Big)
  \Hterm{\hat\nu\hat\rho ij}{}{+\frac{1}{\sqrt{2}}} \hat\epsilon^j~.  \label{eqn:deltaPsi-constraint}
\end{split}
\end{align}
\end{subequations}
Thus, (\ref{eqn:gauge-fixing-constraints}) is solved for
$\hat\epsilon\equiv 0$, $\lambda\equiv 0$ and $\hat X^{\hat\mu}=\left(X^\mu(x),0\right)$,
$\hat\Sigma^{\hat a}_{\ \,\hat c}=\delta^{\hat a}_{\ \,a}\delta^{\ \,c}_{\hat c}\,\Sigma^a_{\ \,c}(x)$,
$\hat\tau^I=\tau^I(x)$ 
and $\hat\sigma=\sigma(x)$,
acting as 
four-dimensional diffeomorphisms $\delta_X$, local Lorentz transformations $\delta_\Sigma$ and
SU(2)$\otimes$U(1) gauge transformations $\delta_{\tau^I}$, $\delta_\sigma$, respectively,  on the boundary fields.

Furthermore, consider 
$\hat\delta_\text{w}{:=}
      \delta_{\hat X_\text{w}}
      {+}\delta_{\hat\epsilon_\text{w}}
      {+}\delta_{\hat\Sigma_\text{w}}
      {+}\delta_{\hat\sigma_\text{w}}
      {+}\delta_{\hat\tau_\text{w}^I}$,
with nonzero $\hat X^r=r\lambda$ accompanied by
$\hat\epsilon_{\text{w}i}=\mathcal O(r^{3/2})$,
by
$\hat X_\text{w}^\mu$,
$\hat\sigma_\text{w}$, $\hat\tau_\text{w}^I$ of $\mathcal O(r^2)$ 
and by $\hat\Sigma_{\text{w}\,b}^{a}=0$, $\hat\Sigma_{\text{w}\,\underline r}^{a}=\mathcal O(r)$
to solve (\ref{eqn:constraints1}), (\ref{eqn:constraints2}).
All transformations preserve the boundary fields,
except for $\delta_{\hat X_\text{w}}$ which acts as a Weyl rescaling.
The Weyl weights of the boundary fields are fixed by the scaling
of the bulk fields from which they are defined, e.g.~for 
$\phi:=\lim_{r\rightarrow 0}r^\alpha\hat\phi$ 
we have 
$\delta_\text{w}\phi:=\lim_{r\rightarrow 0}r^\alpha\hat\delta_\text{w}\hat\phi=-\alpha\lambda(x)\phi$.

Finally, we set $\lambda\equiv 0$ and consider
non-vanishing $\hat\epsilon_i$  solving (\ref{eqn:deltaPsi-constraint}).
Similarly to the mass terms in the spinor field equations,
the $T_{ij}$-term in (\ref{eqn:deltaPsi-constraint}) affects a splitting of the chiral components
when solving the leading order in $r$.
We find the two independent solutions 
$\hat\epsilon_{i_+}=r^{-1/2}\epsilon_{i_+}^\text{L}+o(r^{1/2})$ and
$\hat\epsilon_{i_+}=r^{1/2}\epsilon_{i_+}^\text{R}+o(r^{1/2})$
with $\epsilon_{i_+}^\text{L/R}\vert^{}_{r=0}$ finite.  
$\hat X^\mu$, $\hat\sigma$ and $\hat\tau^I$ of $\mathcal O(r^2)$ and
$\hat\Sigma^{a}_{\ \,\underline r}=\mathcal O(r)$ are fixed by solving
the remaining constraints,
such that
$\delta_{\hat X,\,\hat\Sigma,\,\hat\sigma,\,\hat\tau^I}$ transform the subleading modes of the bulk fields only. 
On the boundary fields we thus have a purely fermionic transformation $\hat\delta_{\hat\epsilon}$.

We define 
$\zeta_{i_+}{:=}\,\epsilon_{i_+}^\text{L}(x,0)$,
$\zeta^{i_+}{:=}\,\epsilon^{\text{R}i_+}(x,0)$,
such that $\zeta^{i_+}$ is related to $\zeta_{i_+}$ by the symplectic Majorana condition,
and similarly $\eta_{i_+}{:=}\,\epsilon_{i_+}^\text{R}(x,0)$, $\eta^{i_+}{:=}\,\epsilon^{\text{L}i_+}(x,0)$.
To leading order in the fermionic fields the $\zeta$-transformations of the boundary fields are
\begin{flalign}\label{eqn:delta-zeta-transformation}
\begin{split}
 \delta_{\zeta}e_\mu^{a} &= i\bar\psi_\mu^{\text{L}i_+}\gamma^a\zeta_{i_+}+\text{c.c.}~,
  \qquad
 \delta_{\zeta}\psi_{\mu i_+}^\text{L} = 
  D_\mu \zeta_{i_+}-\frac{1}{4}\,\gamma\cdot C^-_{i_+j_+}\gamma_\mu\zeta^{j_+}~,
 \\
 \delta_{\zeta} A_{\mu}^I &= \frac{1}{\sqrt{2}}i
                           \Big(\bar\Phi_{\mu}^{\text{R}i_+}\zeta_{j_+} - \frac{1}{\sqrt{3}}\bar\chi^{\text{L}i_+}\gamma_\mu\zeta_{j_+}\Big)
			   \left(\Gamma^I\right)_{i_+}^{\ \ \,j_+}+\text{c.c.}~,
 \\
  \delta_\zeta a_\mu &= \frac{1}{2}i\Big(\bar\Phi_\mu^{\text{R}i_+}\zeta_{i_+}+\frac{2}{\sqrt{3}}\bar\chi^{\text{L}i_+}\gamma_\mu\zeta_{i_+}\Big)+\text{c.c.}~,
  \qquad
  \delta_\zeta\varphi=\frac{1}{\sqrt{2}}i\bar\chi^{\text{R}i_+}\zeta_{i_+}+\text{c.c.}~,
 \\
 \delta_\zeta\chi_{i_+}^\text{L} &= 
      -\frac{1}{\sqrt{2}}i\varphi\zeta_{i_+}
      +\frac{1}{2\sqrt{6}}\gamma^{\mu\nu}\left( 
              F^I_{\mu\nu}\left(\Gamma_I\right)_{i_+}^{\ \ \,j_+}
              -\sqrt{2}f_{\mu\nu}\delta_{i_+}^{\ j_+}
        \right)\zeta_{j_+}
      -\frac{1}{\sqrt{3}}i \gamma^{\mu}C_{\mu r\,i_+ j_+}\zeta^{j_+}~,
  \\ 
  \delta_\zeta C^-_{ab} &=  
      2i\left(\Gamma_4\right)^{i_+ j_+} 
      \Big(\bar\zeta_{i_+}\hat{ R}_{ab\,j_+}(Q)
	+\frac{1}{4}\eta_{ac}\bar\psi^{\mu \text{R}}_{i_+}\gamma^{[\nu}\gamma_{b\mu}\gamma^{c]}\delta_\zeta \psi_{\nu j_+}^\text{L}
      \Big)~,
\end{split}
\end{flalign}
where $\hat R_{\mu\nu\,i_+}(Q):= R_{\mu\nu\,i_+}(Q)-\frac{1}{2\sqrt{3}}i\gamma_{\mu\nu}\chi^\text{L}_{i_+}$.
These correspond to \N{2} (Q-)supersymmetry transformations of the boundary fields.
The $\eta$-transformations are given by
\begin{align}\label{eqn:delta-eta-transformation}
\begin{split}
 \delta_{\eta}e_\mu^{a} &=0~, 
 \qquad
 \delta_{\eta}\psi_{\mu i_+}^\text{L} = -i\gamma_\mu\eta_{i_+}~,\qquad
 \delta_\eta a_\mu = \frac{1}{2}i\bar\psi_\mu^{\text{L}i_+}\eta_{i_+}+\text{c.c.}~,
 \\
 \delta_\eta C^-_{ab}&=
          \frac{1}{2}i \left(\Gamma_4\right)^{i_+ j_+}\eta_{ac}\bar\psi^{\mu \text{R}}_{i_+}\gamma^{[\nu}\gamma_{b\mu}\gamma^{c]}\delta_\eta \psi_{\nu j_+}^\text{L}~,  
 \qquad
 \delta_\eta\varphi = 0~,
 \\
 \delta_\eta\chi_{i_+}^\text{L} &= -\frac{1}{2\sqrt{3}}\,\gamma\cdot C^{-}_{i_+ j_+}\,\eta^{j_+}~,
 \qquad
 \delta_{\eta}A_\mu^I = \frac{1}{\sqrt{2}}i\bar\psi_{\mu}^{\text{L}i_+}\eta_{j_+}\left(\Gamma^I\right)_{i_+}^{\ \ \,j_+}+\text{c.c.}~,
 \end{split}
\end{align}
and correspond to special conformal (S-)supersymmetry or super-Weyl transformations.
The constrained field components $\Phi_{\mu i_+}^\text{R}$, $C_{\mu\nu}^+$ and $C_{\mu r}$ 
are given by (\ref{eqn:gravitino-subdominant}), (\ref{eqn:Cplus}) and 
(\ref{eqn:Cmur}), respectively, and the covariant derivative by (\ref{eqn:covariant-derivative-4d}).
With $\chi_{i_+}^\text{R}$ as given in (\ref{eqn:chi-subdominant}) the transformation of the scalar field may be rewritten as
\begin{align}\label{eqn:delta-phi-2}
\delta_\zeta\varphi=
  \frac{1}{\sqrt{2}}\bar{\zeta}^{i_+}\gamma^\mu\Big(D_\mu-\delta_\zeta(\psi_{\mu})-\delta_\eta(\Phi_{\mu})\Big)\chi_{i_+}^\text{L}
 +\text{c.c.}~,
\end{align}
where $\delta_\zeta(\psi_{\mu})$ denotes a field-dependent $\zeta$-supersymmetry transformation
with parameter $\zeta_{i_+}\!=\psi_{\mu i_+}^\text{L}$,
and analogously for $\delta_\eta(\Phi_{\mu})$ with $\eta_{i_+}\!=\Phi_{\mu i_+}^\text{R}$.

The commutators of Q- and S-supersymmetries can be derived from (\ref{eqn:QQ-commutator})
and we find
\begin{subequations}
\begin{align}
 \left[\delta_{\zeta_2},\delta_{\zeta_1}\right]&=
  \delta_{X_{\zeta}}
  +\delta_{\Sigma}\big(X_{\zeta}^\mu\omega_\mu^{\hphantom{\mu}ab}\big)+\delta_{\Sigma}\big(2i\bar\zeta_1^{i_+}\zeta_2^{j_+}C^{-\,ab}_{\ \ \ \ \ i_+ j_+}\!+\text{c.c.}\big)
  +\delta_{\sigma_\zeta} 
  +\delta_{\tau_\zeta^I}~, \label{eqn:boundary-symmetry-commutator1}
 \\
 \left[\delta_\eta,\delta_\zeta\right]&=\delta_\text{Weyl}\big(\bar\zeta^{i_+}\eta_{i_+}\!+\text{c.c.}\big)
  +\delta_{\Sigma}\big(-\bar\zeta^{i_+}\gamma^{ab}\eta_{i_+}\!+\text{c.c.}\big)
  +\delta_{\sigma_{\eta\zeta}}\label{eqn:boundary-symmetry-commutator2}
  +\delta_{\tau^I_{\eta\zeta}}~,
  \\
 \left[\delta_{\eta_2},\delta_{\eta_1}\right]&=0~,
\end{align}
\end{subequations}
where in (\ref{eqn:boundary-symmetry-commutator1}) the diffeomorphism is
$X^\mu_{\zeta}=-i\bar\zeta_1^{i_+}\gamma^\mu\zeta_{2 i_+}\!+\text{c.c.}$
and the gauge transformations are $\sigma_\zeta=X_{\zeta}^\mu a_\mu$, $\tau_\zeta^I=X_{\zeta}^\mu A_\mu^I$.
The gauge transformations in (\ref{eqn:boundary-symmetry-commutator2}) are 
$\sigma_{\eta\zeta}=\frac{1}{2}i\bar\zeta^{i_+}\eta_{i_+}\!+\text{c.c.}$
and
$\tau^I_{\eta\zeta}=\frac{1}{\sqrt{2}}i\left(\Gamma_I\right)_{i_+}^{\ \ j_+}\bar\zeta^{i_+}\eta_{j_+}\!+\text{c.c.}$.

\begin{table}[htb]
\centering
  \begin{tabular}{l*{6}{c}}
    \toprule
	&  $e_\mu^a$  & $\psi_{\mu i_+}^L$ & $a_\mu$, $A_\mu^I$ &  $\chi_{i_+}^L$ & $C^-_{\mu\nu}$ & $\varphi$\\ 
    \cmidrule(r){2-7}
    $w$\ \ \  & $-1$ & $-\frac{1}{2}$ & $0$ & $\frac{3}{2}$ & $-1$ & $2$\\ \noalign{\smallskip}
    $s$ & $2$ & $\frac{3}{2}$ & $1$ & $\frac{1}{2}$ & $1$ & $0$\\ \noalign{\smallskip}
    $n$ & $5$ & $-8$ & $3$ & $-4$ & $6$ & $1$ \\ \noalign{\smallskip}
    $c$ & $0$ & $\frac{1}{2}$ & $0$  & $\frac{1}{2}$ & $1$ & $0$\\
    \bottomrule
  \end{tabular}
  \caption{\label{tbl:boundary-fields} Boundary fields with Weyl weights $w$, spin $s$ and $n$ off-shell degrees of freedom.
           The fermions are SU(2) doublets and $c$ denotes the U(1) charges. }
\end{table}
Altogether, we find the boundary degrees of freedom with properties as given
in Table~\ref{tbl:boundary-fields} and with the fermionic symmetry transformations
(\ref{eqn:delta-zeta-transformation}), (\ref{eqn:delta-eta-transformation}).
The off-shell degrees of freedom are given as the difference of field components and gauge degrees of freedom,
e.g.~for the chiral gravitino we count $16$ components from which $2\cdot 4$ degrees of freedom are removed 
for the chiral $\zeta$ and $\eta$ supersymmetry transformations.
Likewise, of the $16$ vielbein components $4$ degrees of freedom are subtracted for diffeomorphisms, $6$ for local Lorentz
and $1$ for Weyl transformations.
As seen from Table~\ref{tbl:boundary-fields}, 
the total numbers of bosonic and fermionic degrees of freedom, both being $24$, match nicely,
and
the boundary fields fill the \N{2} Weyl multiplet, see \cite{deWit:1979ug,deRoo:1980mm,Fradkin:1985am}.
The bulk SU(2)$\otimes$U(1) gauge symmetry has become the chiral U(2) transformations 
contained in SU(2,\,2$\vert$2) to close the commutator of Q- and S-supersymmetries.

\section{Application: Holographic Weyl anomaly}\label{sec:Weyl-anomaly}

In this section we give an application of 
the previous results using the AdS/CFT conjecture.
As noted in the introduction, solutions
of Romans' theory can be lifted to the 
ten-dimensional IIA/B supergravities and to the 
maximal $d{=}11$ supergravity.
In particular, the AdS$_5$ vacuum lifts to 
AdS$_5${$\times$}S$^5$ in IIB supergravity \cite{Lu:1999bw} and to a solution
describing the near-horizon limit of a semi-localized system of two sets 
of M5-branes in M-theory \cite{Cvetic:2000yp}.
The latter solution can be understood as uplift of a solution in the IIA theory
describing an elliptic brane system with D4 and NS5 branes \cite{Witten:1997sc,Alishahiha:1999ds,Oz:1999qd}.
Thus, the fluctuations around AdS are understood as a 
dual description of a subsector of \N{4} SYM theory via the 
lift to IIB supergravity, and as dual to the \N{2} SCFTs 
on the M5-brane intersection and on the D4 branes via the
lifts to M-theory and IIA supergravity, respectively.

An important result in AdS/CFT is that the appearance of 
a Weyl anomaly in the SCFT in an external supergravity 
background can be understood holographically as 
follows~\cite{Henningson:1998gx,Balasubramanian:1999re,deHaro:2000xn,Imbimbo:1999bj,Bianchi:2001kw}.
In the limit where string theory is appropriately described by supergravity,
the generating functional of the SCFT correlation functions in the 
conformal supergravity background $g_{\mu\nu},\dots$ 
with sources $\delta g_{\mu\nu},\dots$ 
is related to the path integral of the dual supergravity
as a functional of the boundary conditions 
by~\cite{Marolf:2004fy}
\begin{equation}\label{eqn:adscft}
\int\limits_{[t_-,t_+]}\! \mathcal D\hat g \Big\vert^{}_{r^2\hat g\vert^{}_{\partial X}=g+\delta g} \:
 e^{iS_\text{sugra}[\hat g]} \:
 \langle \hat \beta \vert \hat g,t_+\rangle
 \langle \hat g,t_-\vert\hat\alpha\rangle
  \:=\: \langle\beta\vert T e^{ i \int_{\partial X} \frac{1}{2}\delta g^{\mu\nu}T_{\mu\nu}}\vert\alpha\rangle^{}_\text{SCFT}~.
\end{equation}
The remaining supergravity fields and boundary conditions on the left hand side and the remaining
SCFT operators and sources on the right hand side are implicit.
In the limit where the bulk supergravity becomes classical,
the path integral reduces to the integrand evaluated on 
the solution of the classical field 
equations\footnote{Which involve also boundary conditions at $t_\pm$.
However, for the calculation of the anomaly only the boundary conditions on $\partial X$ are relevant,
since it does not depend on the choice of SCFT state.}.

The on-shell supergravity action, however, is divergent and has to be regularized, 
e.g.\ by introducing an IR cutoff on the radial coordinate $r\geq\epsilon$. 
This reflects the need for regularization of UV divergences on
the CFT side.
The renormalized supergravity action
\begin{equation}\label{eqn:renormalized-sugra-action}
 S_\text{sugra}^\text{ren}=\lim_{\epsilon\rightarrow 0} \big(
S_{\text{sugra},\epsilon}+S_\text{GHY}+S_\text{ct}+S_\text{ct}^{\log}\big)~,
\qquad S_{\text{sugra},\epsilon}=\int_{r\geq\epsilon} d^5x\, \mathcal L_\text{sugra}~.
\end{equation}
is then constructed from the regularized action $S_{\text{sugra},\epsilon}$,
the Gibbons-Hawking-York term $S_\text{GHY}$ for a well-defined variational principle and the 
counterterm action to render the limit $\epsilon\rightarrow 0$ finite.
It turns out that the $1/\epsilon^k$ divergences can be removed by adding 
\mbox{(bulk-)}covariant boundary terms $S_\text{ct}$, constructed e.g.\ from the induced metric.
In odd dimensions, however, there is also a $\log\epsilon$ divergence, 
the counterterm for which explicitly depends on (the coordinate of) the cutoff $\epsilon$.
Due to this explicit cutoff dependence the renormalization breaks invariance under those
bulk diffeomorphisms that act as Weyl rescalings on the boundary.
Applying such a diffeomorphism  inducing a Weyl transformation on the boundary,
as discussed in Section \ref{sec:boundary-symmetries}, yields the anomalous boundary Ward identity corresponding to Weyl invariance.

The anomalous trace of the energy-momentum 
tensor in pure-metric backgrounds has 
been calculated in \cite{Henningson:1998gx} from pure gravity in the bulk.
The extension to dilaton gravity can be found in \cite{Nojiri:1998dh} and higher-order curvature terms in the bulk
arising from higher orders in the effective string-theory action are discussed in \cite{Anselmi:1998zb,Nojiri:1999mh,Blau:1999vz}.
We now study the Weyl anomaly of the SCFTs dual to Romans' theory in generic bosonic \N{2} conformal supergravity backgrounds.
To this end we truncate the five-dimensional \N{4} supergravity to its bosonic sector,
which is consistent because the fermionic field equations are solved trivially
by $\hat\psi_{\hat\mu i}\equiv\hat\chi_i\equiv 0$. 

The bosonic part of the \N{2} Weyl multiplet of boundary fields as determined in the previous section
is given by $\big(e_\mu^a, A_\mu^I, a_\mu, C_{\mu\nu}^-, \varphi\big)$,
and we label the bosonic part of the dual multiplet of SCFT currents by 
$\big(T_\mu^a, J_\mu^{I}, j_\mu, L_{\mu\nu},\phi\big)$.
$T_\mu^a$, $J_\mu^{I}$ and $j_\mu$ are the classically conserved currents\footnote{The dual theory has SU(2)$\otimes$U(1) R-symmetry.} 
and $L_{\mu\nu}$, $\phi$ complete the bosonic part of the supermultiplet.
Equation (\ref{eqn:adscft}) then yields
\begin{align}\label{eqn:adscft-one-point}
\begin{split}
 \delta S_\text{sugra}^\text{ren} &= 
\int_{r=0} d^4x\, e \,\Big(\,
\delta e_{a}^\mu \, \langle T_{\mu}^a \rangle
+ \delta a_\mu \,\langle j^\mu \rangle
+\delta A_\mu^I \,\langle J^{\mu I} \rangle
  + \delta C_{\mu\nu}^{-} \,\langle L^{\mu\nu} \rangle + \delta \varphi\,\langle\phi\rangle
\,\Big)~.
\end{split}
\end{align}
We choose the variations of the boundary conditions such that they correspond to a Weyl transformation,
$\delta e_{\mu}^a=-\lambda e_\mu^a$ and likewise for the remaining fields.
Extending them into the bulk to a diffeomorphism as discussed in Section~\ref{sec:boundary-symmetries},
generated by the vector field
$(X^\mu,\lambda r)$ with $\partial_r X^\mu=r g^{\mu\rho}\partial_\rho\lambda$ and $X^\mu\vert_{r=0}=0$, yields the
anomalous Ward identity
\begin{flalign} \label{eqn:Ward-id-anom}
 \langle T_\mu^{\hphantom{\mu}\mu}\rangle - C_{\mu\nu}^{-}\langle L^{\mu\nu}\rangle + 2\varphi\langle\phi\rangle &= \mathcal A~,
\qquad \mathcal A:=\lim\limits_{\epsilon\rightarrow 0}\,\frac{1}{e}\frac{\delta}{\delta\lambda} S_\text{ct}^\text{log}~.
\end{flalign}
In the remaining part of this section we will determine $\mathcal A$. To this end we have to solve
the nonlinear field equations of the various fields as asymptotic series in a vicinity of the boundary,
which then allows us to determine the divergences of the on-shell action and the required counterterms.

\subsection{On-shell bulk fields as asymptotic series}\label{sec:Weyl-anomaly-on-shell-fields}
We now determine the required subleading modes of the bulk fields from their field equations.
For the matter fields the equations have been given in Section \ref{sec:boundary-fields} and 
Einstein's equations for the bosonic sector read\footnote{Our 
conventions are 
$\hat \calR_{\hat \mu}^{\hphantom{\hat\mu}\hat a}(\hat\omega)
=
\hat e_{\hat b}^{\hat\nu}\hat \calR_{\hat \mu\hat\nu}^{\hphantom{\hat\mu\hat\nu}\hat a\hat b}(\hat\omega)$
and
$\hat \calR_{\hat\mu\hat\nu}(\hat\omega)
=
\hat e_{\hat\nu\hat b}\hat \calR_{\hat \mu}^{\hphantom{\hat\mu}\hat a}(\hat\omega)$.
}
\begin{flalign}\label{eqn:Einstein-bosonic}
\begin{split}
  \hat\calR_{\hat\mu\hat\nu}(\hat\omega)
    =\,&
  \frac{4}{3}P(\hat\varphi)\hat g_{\hat\mu\hat\nu} + 2 \hat D_{\hat\mu}\hat\varphi\hat D_{\hat\nu}\hat\varphi
  -\xi^{-4}\big(
              2\hat f_{\hat\mu\hat\rho}\hat f_{\hat\nu}^{\hphantom{\hat\nu}\hat\rho}
              -\frac{1}{3}\hat g_{\hat\mu\hat\nu}\hat f^{\hat\rho\hat\sigma}\hat f_{\hat\rho\hat\sigma}
           \big)
\\ &
   -\xi^2\Big(
         2\hat B_{\hat\mu\hat\rho}^{\ \alpha} \hat B_{\hat\nu}^{\hphantom{\hat\nu}\hat\rho\alpha}
         +2\hat F_{\hat\mu\hat\rho}^{\ I} \hat F_{\hat\nu}^{\hphantom{\hat\nu}\hat\rho I}
         -\frac{1}{3}\hat g_{\hat\mu\hat\nu}
              \big(
                   \hat B^{\hat\rho\hat\sigma\alpha} \hat B_{\hat\rho\hat\sigma}^{\ \alpha}
                   +\hat F^{\hat\rho\hat\sigma I} \hat F_{\hat\rho\hat\sigma}^{\ I}
              \big)
        \Big)~.
\end{split}
\end{flalign}
The coupled system of equations can be solved order by order in an expansion around the asymptotic boundary.
The leading order has been discussed in Sections \ref{sec:boundary-fields} and  \ref{sec:nonlinear-theory}.
To consistently solve the Dirichlet problem $\log r$ terms have to be included in the expansions,
which yields the asymptotic forms
\begin{flalign}
\begin{split}
  g_{\mu\nu}(x,r) &= g_{\mu\nu}^\sbr{0}+r^2 g_{\mu\nu}^\sbr{2}+r^3g_{\mu\nu}^\sbr{3}
 +r^4(\log r)^2 h_{\mu\nu}^\sbr{0} +r^4\log r\, h_{\mu\nu}^\sbr{1}+r^4 \tilde g_{\mu\nu}^\sbr{4}+o(r^4)~,
\\
\hat a_\mu(x,r)&=a^\sbr{0}_\mu+ o(r)~,
\\
\hat A_\mu(x,r)&=A^{I\sbr{0}}_\mu+ o(r)~,
\\
\hat C_{\mu\nu}(x,r)&=r^{-1}C_{\mu\nu}^{-\sbr{0}}+r\log r\, C_{\mu\nu}^{+\sbr{1}}+r\, \tilde C_{\mu\nu}^{\sbr{2}}+o(r)~,
\quad
\hat C_{\mu r}(x,r) = C_{\mu r}^\sbr{0}+\mathcal O(r^2\log r)~,
\\
\hat\varphi(x,r)&=r^2\log r\, \varphi^\sbr{0}+r^2\tilde\varphi^\sbr{1}+o(r^2)~.
\end{split}
\end{flalign}
The leading mode $C_{\mu\nu}^{-\sbr{0}}$ of the tensor field is anti-selfdual
and the $r\log r$ term $C_{\mu\nu}^{+\sbr{1}}$ selfdual.
Note the additional $r^4(\log r)^2$ term in the metric expansion as compared to the pure-gravity case.
This is necessary due to the $\hat\varphi^2$ 
term in (\ref{eqn:Einstein-bosonic}).
Due to the additional $\log$-terms and the fact that $h_{\mu\nu}^\sbr{1}$ is not traceless (as will be seen below),
the bulk-covariant counterterms cancelling the $1/\epsilon^k$-divergences do contribute additional $\log$-divergences 
(in contrast to the pure-gravity case)
and we have to determine them first.
The 2$^\text{nd}$-order field equations fix the bulk fields in terms of two sets of boundary data.
Namely, the bulk metric $\hat g$ is fixed from the boundary metric $g^\sbr{0}$ 
and the traceless and divergence-free part of $\tilde g^\sbr{4}$, 
the two-form field $\hat C$ is determined by specifying the anti-selfdual boundary field
$C^{-\sbr{0}}$ and the selfdual part of $\tilde C^\sbr{2}$,
and the Dirichlet data for $\hat \varphi$ is given by $\varphi^\sbr{0}$ and $\tilde \varphi^\sbr{1}$.
Thus, only the leading modes of the on-shell bulk fields are fixed in terms of the boundary fields alone.
The second set of boundary data is linked to the choice of SCFT states.

To determine $g^\sbr{2}_{\mu\nu}$ we need the $\mu\nu$-components of the Ricci tensor for the metric (\ref{eqn:Fefferman-Graham}).
With a prime denoting differentiation with respect to $r$ and 
$R_{\mu\nu}(\omega)$ being the curvature of the four-dimensional spin connection $\omega_{\mu ab}$
they read
\begin{flalign}
\begin{split}
 \hat R_{\mu\nu}(\hat\omega) =\,& 
    R_{\mu\nu}(\omega) + \frac{4}{r^2}g_{\mu\nu} - \frac{3}{2r}g^\prime_{\mu\nu}
    +\big(\frac{1}{4}g^\prime_{\mu\nu}-\frac{1}{2r}g_{\mu\nu}\big)\tr g^{-1}g^\prime
    +\frac{1}{2}g^{\prime\prime}_{\mu\nu}-\frac{1}{2}g^\prime_{\mu\rho} g^{\rho\sigma}g^\prime_{\sigma\nu}~.
\end{split}
\end{flalign}
Solving the $\mu\nu$-components of (\ref{eqn:Einstein-bosonic}) at $\mathcal O(r^{-1})$
shows that there is no contribution to $g_{\mu\nu}(x,r)$ linear in $r$.
Solving at $\mathcal O(r^0)$  shows
\begin{align}
 g_{\mu\nu}^\sbr{2} &=
       \frac{1}{2}\Big(
          \calR^\sbr{0}_{\mu\nu}(\omega)
          -\frac{1}{6}\calR^\sbr{0}(\omega)g_{\mu\nu}^\sbr{0}
          +4\,\overline{C_{\mu\rho}^{-\sbr{0}}}\, C_{\hphantom{-\sbr{0}}\nu}^{-\sbr{0}\hphantom{\nu}\rho}
      \Big)~.
\end{align}
Note that the last term is real due to the anti-selfduality of $C_{\mu\nu}^{-\sbr{0}}$.
For the gauge fields we find from (\ref{eqn:eom-amu}) and (\ref{eqn:eom-AImu}) that the first subleading modes are $o(r)$.
Equation (\ref{eqn:Ctensor-eom}) yields
\begin{equation}
C_{\mu r}^\sbr{0}=\frac{1}{2}i e^{\sbr{0}-1}\epsilon_{\mu}^{\hphantom{\mu}\rho\sigma\tau}D_\rho C_{\sigma\tau}^{-\sbr{0}}~,
\qquad
 C_{\mu\nu}^{+\sbr{1}}=\big(\mathds{1}+i\star^\sbr{0}\!\big)\big( g_{\,[\mu}^{\sbr{2}\,\rho}C_{\nu]\rho}^{-\sbr{0}}
   -  D_{[\mu} C_{\nu]r}^\sbr{0}\big)~.
\end{equation}
For the on-shell action we also need the expansion of the vielbein determinant
\begin{equation}\label{eqn:vielbein-det-asymptotic}
 e=e^\sbr{0}\Big(
    1+\frac{1}{2}r^2t^\sbr{2}+\frac{1}{2}r^4(\log r)^2 u^\sbr{0}+\frac{1}{2}r^4\log r\,u^\sbr{1}
    +\frac{1}{2}r^4\big(t^\sbr{4}+\frac{1}{4}(t^\sbr{2})^2-\frac{1}{2}t^\sbr{2,2}\big)
   \Big)+o(r^4)~,
\end{equation}
where $t^\sbr{n}:=\tr g^{\sbr{0}\,-1}g^\sbr{n}$, $u^\sbr{n}:=\tr g^{\sbr{0}-1}h^\sbr{n}$
and $t^\sbr{2,2}:=\tr g^{\sbr{0}\,-1}g^\sbr{2} g^{\sbr{0}\,-1}g^\sbr{2}$.
These traces can be determined from the $rr$-components of (\ref{eqn:Einstein-bosonic})
with
\begin{equation}
 \hat\calR_{rr}(\hat\omega)=
   -\frac{4}{r^2}+\frac{1}{2r}\tr g^{-1}g^\prime-\frac{1}{2}\tr g^{-1}g^{\prime\prime}
   +\frac{1}{4}\tr g^{-1}g^\prime g^{-1} g^\prime~.
\end{equation}
For notational convenience we define
$\hat{\overline C}_{\rho\sigma}\hat C^{\rho\sigma}=:r^4\log r\, c^\sbr{0}+r^4 c^\sbr{1}+o(r^4)$.
The leading term, which would be $\mathcal O(r^2)$, vanishes due to the anti-selfduality
of $C_{\rho\sigma}^{-\sbr{0}}$.
With
$b^\sbr{0}:=r^{-2}\hat B_{r\rho}^{\ \alpha}\hat B_{r}^{\hphantom{r}\rho\alpha}\vert_{r=0}$
we find
$t^\sbr{3}=0$ and 
\begin{align}
\begin{split}
 u^\sbr{0} &=-\frac{4}{3}{\varphi^\sbr{0}}^2~,
\qquad
 u^\sbr{1} = -\frac{8}{3}\varphi^\sbr{0}\tilde\varphi^\sbr{1}
             +\frac{1}{6} c^\sbr{0}~,
\\
t^\sbr{2,2}-4t^\sbr{4}&=
  \frac{16}{3} { {}\tilde\varphi^\sbr{1} }^2 +\frac{2}{3}{\varphi^\sbr{0}}^2
  -\frac{1}{3}\big(F^{\sbr{0}\rho\sigma I}F_{\rho\sigma}^{\sbr{0}I}+f^{\sbr{0}\rho\sigma}f_{\rho\sigma}^\sbr{0}\big)
  -\frac{4}{3}b^\sbr{0}
  -\frac{2}{3}c^\sbr{1}+\frac{1}{2}c^\sbr{0}~.
\end{split}
\end{align}
Note the dependences on $c^\sbr{1}$ and $\tilde\varphi^\sbr{1}$ which are not fixed by the near-boundary analysis.

\subsection{Holographic renormalization}\label{sec:Weyl-anomaly-holographic-renormalization}
Having calculated the necessary terms in the asymptotic expansions of the bulk fields we now determine
the divergences of the on-shell action and the necessary counterterms.
Using (\ref{eqn:Einstein-bosonic}) and (\ref{eqn:Ctensor-eom})
the Lagrangian (\ref{eqn:N4-Lagrangian1}) truncated to the bosonic sector
reads
\begin{align}\label{eqn:on-shell-Lagrangian-bosonic}
 \mathcal L_\text{on-shell}&=
   -2\hat e -\frac{4}{3}\hat e \hat\varphi^2
   +\frac{1}{12}\hat e \xi^2\hat B^{\hat\mu\hat\nu\alpha}\hat B_{\hat\mu\hat\nu}^{\ \alpha}
   -\frac{1}{6}\hat e \big( \hat F^{\hat\mu\hat\nu I}\hat F_{\hat\mu\hat\nu}^{\ I}
   +\hat f^{\hat\mu\hat\nu}\hat f_{\hat\mu\hat\nu}\big)
   +\mathcal O(r^0)~.
\end{align}
Na\"ively, one may expect terms of order $r^{-1}(\log r)^2$ and $r^{-1}\log r$ in $\mathcal L_\text{on-shell}$,
e.g.\ due to the scalar and tensor field terms.
This potentially leads to $(\log\epsilon)^3$ and $(\log\epsilon)^2$ divergences in the on-shell action.
However, it turns out that the contributions from $\hat e$ to these terms cancel the others, such that only terms
proportional to $r^{-n}$ with $n=5,3,1$ and $\mathcal O(r^0)$ are nonvanishing in $\mathcal L_\text{on-shell}$.
As may be verified with the expansions of the previous section, $S_{\text{sugra},\epsilon}+S_\text{GHY}+S_\text{ct}$ with
\begin{align}\label{eqn:S-ct}
\begin{split}
 S_\text{GHY} = \frac{1}{2}\int\limits_{r=\epsilon}d^4x\, \hat e^\ast \hat K~,
 \qquad
S_\text{ct}= \int\limits_{r=\epsilon}d^4x\, \hat e^\ast
      \Big( 
          -\frac{3}{2} +\frac{1}{8}\mathcal R^\ast(\omega)-\hat\varphi^2 + 
           \alpha\, \hat{\overline{C}}{}^{\ast\mu\nu}\hat C^{\ast}_{\mu\nu}
      \Big)~,
\end{split}
\end{align}
only has a logarithmic divergence in the limit $\epsilon\rightarrow 0$, i.e.\ all $1/\epsilon^k$ divergences are cancelled.
The $\ast$ denotes induced quantities on the boundary, e.g.\ the pullback of the vielbein and the two-form field $\hat C$,
and indices are contracted with the induced vielbein and metric.
$\hat K:=\hat e^{\hat\mu}_{\hat a}\hat K_{\hat\mu}^{\hphantom{\hat\mu}\hat a}$ 
is the trace of the extrinsic curvature of the boundary\footnote{%
The extrinsic curvature is defined as
$\hat K_{\hat\mu\hat\nu}:=P_{\hat\mu}^{\hphantom{\hat\mu}\hat\rho} P_{\hat\nu}^{\hphantom{\hat\nu}\hat\sigma}\hat\nabla_{\hat\rho}\hat n_{\hat\sigma}$ 
with the projector
$P_{\hat\mu}^{\hphantom{\hat\mu}\hat\nu}=\delta_{\hat\mu}^{\hphantom{\hat\mu}\hat\nu}-\hat n_{\hat\mu}\hat n^{\hat\nu}/\hat g(\hat n,\hat n)$
and the outward-pointing unit normal vector field $\hat n^{\hat\mu}=\hat e^{\hat\mu\underline{r}}$.
Using the vielbein postulate this yields $\hat K_{\hat\mu}^{\hphantom{\hat\mu}\hat a}=\hat\omega_{\hat\mu}^{\hphantom{\hat\mu}\hat a\underline{r}}$.
}, 
$\hat K_{\hat\mu}^{\hphantom{\hat\mu}\hat a}=\hat\omega_{\hat\mu}^{\hphantom{\hat\mu}\hat a\underline{r}}$.
We note that also $S_\text{GHY}$ and $S_\text{ct}$ separately only have
power-law and $\log\epsilon$ divergences, e.g.\
the $\hat\varphi^2$ term in $S_\text{ct}$ cancels the $(\log\epsilon)^2$ divergence
in the cosmological-constant term $-\frac{3}{2}\hat e^\ast$.
Thus, the only remaining divergence is $\log\epsilon$, which is consistent with the expectation that the Weyl anomaly of
the dual theory is exhausted at one-loop\footnote{%
It shares a multiplet with the chiral anomaly which receives no contributions beyond one-loop order.
}.
A slight subtlety arises for the $\hat e^\ast\hat{\overline{C}}{}^{\ast\mu\nu}\hat C^{\ast}_{\mu\nu}$ term in $S_\text{ct}$.
Since
the leading term vanishes on-shell thanks to the anti-selfduality of $C_{\mu\nu}^{-\sbr{0}}$, 
it only contributes a logarithmic divergence.
However, as it does not explicitly depend on the cutoff and therefore does not contribute to the Weyl anomaly we include it 
in $S_\text{ct}$ with a for now arbitrary numerical coefficient $\alpha$.

The remaining counterterm required to cancel the $\log\epsilon$ divergence depends on $\alpha$ and is given by
\begin{align}\label{eqn:S-ct-log}
\begin{split}
 S_\text{ct}^{\log} =\int_{r=\epsilon}d^4 x\:\hat e^\ast \bigg( &  
  \frac{1}{16}\Big(\mathscr{R}_{\mu\nu}\mathscr{R}^{\mu\nu}-\frac{1}{3}\mathscr{R}^2\Big)\log\epsilon
  -\frac{1}{4}\big(\hat F_{\mu\nu}^{\ast I}\hat F^{\ast \mu\nu I}+\hat f^\ast_{\mu\nu}\hat f^{\ast\mu\nu}\big)\log\epsilon
 \\ &
  -\frac{1}{2}\hat\varphi^2  (\log\epsilon)^{-1}
  -\big(D_a \hat {\overline{C}}{}_{\hphantom{\ast}b}^{\ast \hphantom{b}a}\big)\big(D_c \hat C^{\ast bc}\big) \log\epsilon
 +(1-4\alpha) \mathcal B \log\epsilon\bigg)~,
\end{split}
\end{align}
where we defined the modified curvature 
$\mathscr{R}_{\mu\nu}:=\calR^\ast_{\mu\nu}(\omega)+4\,\hat{\overline{C}}{}^{\,\ast }_{\mu\rho}\, \hat C_{\hphantom{\ast}\nu}^{\ast\hphantom{\nu}\rho}$
and $D_a$ is the covariant derivative with the four-dimensional spin connection $\omega_{\mu ab}$.
The dependence on $\alpha$ is seen in the last term, where 
\begin{align*}
 \mathcal B = 
   \big(D_a \hat {\overline{C}}{}_{\hphantom{\ast}b}^{\ast \hphantom{b}a}\big)\big(D_c \hat C^{\ast bc}\big)
   -\frac{1}{2}\mathscr{R}^{\mu\nu}\,\hat{\overline{C}}{}^{\,\ast}_{\mu\rho} \,\hat C^{\ast\hphantom{\nu}\rho}_{\hphantom{\ast}\nu}
   -\frac{1}{2} D_a D_b\big(\,\hat{\overline{C}}{}^{\ast ac}\hat C^{\ast b}_{\hphantom{\ast b}c}\,\big)~.
\end{align*}
Clearly, the choice of $\alpha$ will affect the Weyl anomaly, so it has to be fixed.
As the renormalized bulk action should yield finite correlation functions
for the boundary theory, 
we calculate the one-point function of the energy-momentum tensor of the
dual Yang-Mills theory.
According to (\ref{eqn:adscft-one-point}) it is given by
\begin{flalign}\label{eqn:Tmunu-onepoint}
 \langle T_\mu^a\rangle
  \,=\,
 \frac{1}{e^\sbr{0}}\frac{\delta S^\text{ren}_\text{sugra}}{\delta e^{\mu\sbr{0}}_{a}}
  \,=\,\lim\limits_{\epsilon\rightarrow 0}\epsilon^{-3}\frac{1}{\hat e^\ast}
       \frac{\delta S^\text{ren}_{\text{sugra},\epsilon}}{\delta \hat e^{\ast\mu}_{\:a}}
  \,=:\,
 \lim\limits_{\epsilon\rightarrow 0}\epsilon^{-3}\, \mathcal T_\mu^a,
\end{flalign}
where $S^\text{ren}_{\text{sugra},\epsilon}$ is the action defined in (\ref{eqn:renormalized-sugra-action}) 
before taking the limit $\epsilon\rightarrow 0$.
$\mathcal T_\mu^a$ is the Brown-York quasilocal energy-momentum tensor \cite{Brown:1992br} of the bulk supergravity with
regularization $r\geq\epsilon$ and supplemented by the counterterms.
We find
\begin{flalign}\label{eqn:calTmunu}
\begin{split}
 \mathcal T_\mu^a \,=\ & \frac{1}{2}\big(\hat K_{\mu}^{\hphantom{\mu}a} -\hat e_{\:\mu}^{\ast a} \hat K\big) 
                        +\frac{3}{2}\hat e_{\:\mu}^{\ast a}
                        + \frac{1}{4}\big(\mathcal R_\mu^{\ast a}(\omega)   -\frac{1}{2}\hat e_{\:\mu}^{\ast a}\mathcal R^\ast(\omega)\big)
   \\ &
                     +2\alpha \big(\hat{\overline{C}}{}^{\ast}_{\mu\nu}\hat C^{\ast a \nu}+\text{c.c.}\big)
                     -\alpha \hat e_{\:\mu}^{\ast a}\, \hat{\overline{C}}{}^{\,\ast}_{\nu\rho}\hat C^{\ast \nu\rho}
                     +\frac{1}{e^\ast}\frac{\delta S_\text{ct}^{\log}}{\delta \hat e^{\ast\mu}_{\:a}}~.
\end{split}
\end{flalign}
Inserting the on-shell expansion of the fields as obtained in Section \ref{sec:Weyl-anomaly-on-shell-fields},
the leading part of $\hat{\overline{C}}{}^{\,\ast}_{\mu\nu}\hat C^{\ast a \nu}$ does not vanish and
contributes at $\mathcal O(\epsilon)$.
Demanding a finite limit in (\ref{eqn:Tmunu-onepoint}) then fixes $\alpha=\frac{1}{4}$.
Similarly, finiteness of $\langle L^{\mu\nu}\rangle$ also requires this choice of $\alpha$.
The reason why finiteness of the one-point functions requires a fixed $\alpha$
while finiteness of the on-shell action does not can be seen as follows.
The vanishing of the leading order of the counterterm
$\hat{\overline{C}}{}^{\ast\mu\nu}\hat C^{\ast}_{\mu\nu}$
due to the anti-selfduality of $C_{\mu\nu}^{-\sbr{0}}$
relies on the contraction of the two-form fields with the metric.
Therefore, finiteness of the action evaluated on solutions of the 
classical field equations does not guarantee finiteness of the 
variations with respect to the metric or the two-form field evaluated on
classical solutions.

Finally, we obtain the anomalous contribution to the Ward identity (\ref{eqn:Ward-id-anom}) from the variation of
(\ref{eqn:S-ct-log}) for $\alpha=\frac{1}{4}$ and find,
with
$\mathscr{R}^\sbr{0}_{\mu\nu}=\calR^\sbr{0}_{\mu\nu}(\omega)+4\,\overline{C^{-\sbr{0}}_{\mu\rho}}\, C_{\hphantom{-\sbr{0}}\nu}^{-\sbr{0}\hphantom{\nu}\rho}$,
\begin{align} \label{eqn:weyl-anomaly}
\begin{split}
 \mathcal A=\,&
  -\frac{1}{16}\Big(\mathscr{R}^\sbr{0}_{\mu\nu}\mathscr{R}^{\sbr{0}\mu\nu}-\frac{1}{3}{\mathscr{R}^\sbr{0}}^2\Big)
  +\,\overline{D_a C_{\hphantom{-\sbr{0}}b}^{-\sbr{0}\hphantom{b}a}}\, D_c C^{-\sbr{0}bc}
 \\ &
  -\frac{1}{2}{\varphi^\sbr{0}}^2
  +\frac{1}{4}\Big(F_{\mu\nu}^{\sbr{0}I}F^{\sbr{0}\mu\nu I}+f^\sbr{0}_{\mu\nu}f^{\sbr{0}\mu\nu}\Big)
  ~.
\end{split}
\end{align}
The curvature-squared part of the first term yields the difference of the squared Weyl tensor
and the four-dimensional Euler density, and
the mixed terms complete the kinetic term of the two-form field to its Weyl-invariant form. 
Note that the anomaly depends on the boundary fields only, 
i.e.\ the dependences on $\tilde\varphi^\sbr{1}$ and $c^\sbr{1}$, which
are not fixed by the near-boundary analysis,  have dropped out.
This is to be expected as the anomaly is a UV effect in the dual theory.
From the dual Yang-Mills theory point of view, the Weyl anomaly of 
\N{4} SYM theory should be given by the Lagrangian of \N{4} conformal supergravity \cite{Liu:1998bu}.
As noted before, the bulk theory discussed here provides a holographic description of
a subsector of that theory and thus the Weyl anomaly should correspond to a subsector of the
\N{4} conformal supergravity Lagrangian.
Comparing the holographic Weyl anomaly (\ref{eqn:weyl-anomaly})
to the construction of four-dimensional extended conformal supergravity in \cite{Bergshoeff:1980is}, 
it indeed matches the bosonic part of the \N{2} conformal supergravity Lagrangian (5.18) of \cite{Bergshoeff:1980is}.
Thus, our result gives further support to the AdS/CFT conjecture.

\section{Conclusion} \label{sec:conclusions}
In this paper we have studied SU(2)$\otimes$U(1) gauged \N{4} supergravity 
on asymptotically-AdS$_5$ backgrounds.
We have constructed the multiplet of fields induced on the conformal boundary
and determined the induced representation of the local bulk symmetries on the
boundary fields.
This has shown that the boundary degrees of freedom, which are  given in Table~\ref{tbl:boundary-fields},
fill the \N{2} Weyl multiplet and that the complete local \N{2} superconformal transformations 
are induced, with Q- and S-supersymmetry transformations given 
in (\ref{eqn:delta-zeta-transformation}), (\ref{eqn:delta-eta-transformation}).

For the constructions we have employed gauge fixings for the bulk symmetries, which were chosen such
that they do not cause a fixing of the symmetries induced on the boundary.
Different gauge fixings are expected to yield the same boundary fields and symmetries, 
possibly gauge fixed and/or with additional gauge degrees of freedom.
An interesting task is to study this in the BRST approach.
Note also that for the cases discussed here the rescaled boundary limit 
of the bulk fields agrees with their rescaled pullback to the boundary.

In the second part we have used these results and the AdS/CFT conjecture to study the 
four-dimensional SCFTs dual to Romans' theory,
e.g.\ the worldvolume theory on the D4-branes of the elliptic brane configuration
studied in \cite{Witten:1997sc,Alishahiha:1999ds,Oz:1999qd}.
For that purpose,
we have carried out the holographic renormalization of the bosonic sector
of the gauged \N{4} supergravity.
As we have seen, the boundary terms  (\ref{eqn:S-ct}), (\ref{eqn:S-ct-log})
ensure finiteness of the action
evaluated on solutions of the classical field equations,
and for $\alpha=\frac{1}{4}$ also of the variations of the action
evaluated on the classical solutions.
In particular, we found a finite SCFT energy-momentum tensor which is obtained
as the rescaled boundary limit of the Brown York energy-momentum tensor of the bulk theory (\ref{eqn:calTmunu}).
The boundary terms (\ref{eqn:S-ct-log}) break part of the bulk diffeomorphism invariance,
which leads to the anomalous contribution (\ref{eqn:weyl-anomaly}) to the boundary Ward identity for 
Weyl invariance~(\ref{eqn:Ward-id-anom}).
Thus, we have obtained the Weyl anomaly for the dual SCFTs in a generic bosonic \N{2} conformal supergravity
background, including the matter field contributions.

An interesting point for further investigation concerns the holographic counterterms.
Remarkably, as shown in \cite{Grumiller:2009dx}, for lower-dimensional theories
the holographic counterterms coincide with the boundary terms required by supersymmetry 
in the presence of a boundary.
It would certainly be interesting to see whether demanding supersymmetry is sufficient to reproduce
the boundary terms obtained here.
Furthermore, the renormalized action and Brown-York energy-momentum tensor (\ref{eqn:calTmunu})
may be useful for characterizing solutions of Romans' theory involving matter fields,
e.g.\ for the solutions with non-Abelian gauge fields discussed in \cite{Radu:2006va}.

Another  field for further research is in a somewhat different direction.
A duality relation reminiscent of the AdS/CFT correspondence has been formulated 
and proven in \cite{Rehren:1999jn, Rehren:2000tp} in the context of algebraic QFT. 
Although it is unclear whether the bulk theory considered here
can be fit into the framework of algebraic QFT, we may still try to interpret the
results of the first part on the asymptotic structure in that context%
\footnote{
\cite{Rehren:1999jn, Rehren:2000tp} relies on the precise properties of
AdS space, so we may regard the bulk theory as expanded around an 
AdS background for that purpose.
}.
While the physical interpretation of the boundary theory in \cite{Rehren:1999jn, Rehren:2000tp} is not 
immediately clear, the constructions in \cite{Bertola:2000pp, Rehren:2004yu},
where the boundary Wightman field is constructed as boundary limit of the rescaled AdS Wightman field,
suggest that the boundary fields constructed here indeed constitute the field content of the
boundary theory.
This may also be understood in the context of \cite{Compere:2008us}, 
where it was shown that, replacing the Dirichlet boundary conditions employed 
in the AdS/CFT correspondence by Neumann or mixed boundary conditions,
the boundary metric can be promoted to a dynamical field.
An interesting task left for the future is to combine our results with the
appropriate boundary conditions to construct a dynamical conformal supergravity on the boundary.

\section*{Acknowledgements}
CFU thanks Alexander Schenkel and Claudio Dappiaggi for useful discussions.
CFU is supported by the German National Academic Foundation 
(Studienstiftung des deutschen Volkes) and by Deutsche
Forschungsgemeinschaft through the Research Training Group GRK\,1147 
\textit{Theoretical Astrophysics and Particle Physics}.

\appendix
\section{Conventions} \label{app:conventions}
In this appendix we give a summary of the conventions for 
the $\mathfrak{usp}(4)$ generators, which agree with those of \cite{Romans:1985ps},
and for the spacetime $\gamma$-matrices. 
All spacetime quantities are five-dimensional, so we omit hats for better readability.
The $\gamma$-matrices are chosen such that
$\gamma_{abcde}=\epsilon_{abcde}$ with $\epsilon_{01234}=1$. 
With the charge conjugation matrix $C$ satisfying
\begin{align}\label{eqn:charge-conjugation-C}
 C\hat\gamma_{\hat\mu} C^{-1}=\hat\gamma_{\hat\mu}^T,\qquad C^T=C^{-1}=-C,\qquad C^\star=C
\end{align}
the supercharges and hence all the spinors satisfy the symplectic Majorana condition
\begin{align}\label{eqn:sympMaj}
 \left(\chi_i\right)^\dagger\gamma_0 =: \bar\chi^i = \left(\chi^i\right)^T C~.
\end{align}
Fermionic fields are by convention anticommuting and complex conjugation changes their order.
Antisymmetrized indices are defined as $X_{[\mu}Y_{\nu]}:=\frac{1}{2}(X_\mu Y_\nu-X_\nu Y_\mu)$.

The  $\mathfrak{usp}(4)$ symplectic metric $\Omega$ and its inverse satisfy 
$\Omega_{ij}\Omega^{jk}=\delta_i^{\ k}$, 
$\Omega^{ij}=\left(\Omega_{ji}\right)^\ast$
and are used to raise and lower spinor indices via
$\epsilon^i=\Omega^{ij}\epsilon_j$ and $\epsilon_i=\Omega_{ij}\epsilon^j$.
The $\mathfrak{so}(5)$ Clifford algebra generators $\Gamma_m$ satisfy
$\left(\Gamma_m\right)_{i}^{\ \,k} \left(\Gamma_n\right)_{k}^{\ \,j}+\left(\Gamma_n\right)_{i}^{\ \,k} \left(\Gamma_m\right)_{k}^{\ \,j} = 2\delta_{mn}\delta_{i}^{\ \,j}$,
which yields canonical Clifford matrices only for these specific index positions.
With the charge conjugation matrix $\Omega$ we have
\begin{align}
 \Omega^{ik}\left(\Gamma_m\right)_k^{\ \,j} =: \left(\Gamma_m\right)^{ij} = -\left(\Gamma_m\right)^{ji}~.
\end{align}
The conjugate is denoted by $\left(\Gamma_{m}\right)_{ij}=\big(\left(\Gamma_{m}\right)^{ij}\big)^\ast$
and
the $\mathfrak{so}(5)$ generators satisfy $\left(\Gamma_{mn}\right)^{ij}=\left(\Gamma_{mn}\right)^{ji}$.
The convention for $\epsilon^{\alpha\beta}$ is $\epsilon_{45}=\epsilon^{45}=1$.

As usual, the Landau notation is defined by
\begin{equation}
 f\stackrel{x\rightarrow x_0}{=}\mathcal O(g) \  :\Longleftrightarrow \  \limsup\limits_{x\rightarrow x_0}|f/g|<\infty~,
\qquad
f\stackrel{x\rightarrow x_0}{=} o(g) \  :\Longleftrightarrow \  \lim\limits_{x\rightarrow x_0}|f/g|=0~.
\end{equation}

\section[Comparison to the literature]{Comparison to \cite{deWit:1979ug, deRoo:1980mm} }
To connect the superconformal transformations 
(\ref{eqn:delta-zeta-transformation}), (\ref{eqn:delta-eta-transformation}), (\ref{eqn:delta-phi-2}) 
to the results obtained in \cite{deWit:1979ug, deRoo:1980mm}
we first redefine the tensor field as  
$\mathcal C_{\mu\nu}:=C_{\mu\nu}-i\left(\Gamma_4\right)^{i_+ j_+}\bar\psi^\text{R}_{[\mu\,i_+}\psi^\text{L}_{\nu]\,j_+}$
such that, to leading order in the fermions,
\begin{align}
  \delta_\zeta \mathcal C^-_{ab} &=        2i\left(\Gamma_4\right)^{i_+ j_+}       \bar\zeta_{i_+}\hat{ R}_{ab\,j_+}(Q)      ~,
\qquad\quad
 \delta_\eta\mathcal C^-_{ab}=0~,
\end{align}
while the transformations of the other fields change by $C^-{\rightarrow}\ \mathcal C^-$ only.
With the field redefinitions
\begin{flalign}
 \psi_{\mu i_+}^\text{L}&=:\frac{\kappa}{\sqrt{2}} \Psi_{\mu \iota}~,
 \qquad
 \Phi_{\mu i_+}^\text{R}+\frac{1}{2\sqrt{3}}\gamma_\mu\chi_{i_+}^\text{L}=:\frac{\kappa}{\sqrt{8}}\varPhi_{\mu\iota}~,
 \qquad
 \hat R_{\mu\nu\,i_+}(Q)=: \frac{\kappa}{\sqrt{8}} \hat R^\prime_{\mu\nu\,\iota}(Q)~,\nonumber\\
 \zeta_{i_+}&=:\sqrt{2}\zeta^\prime_\iota~,
 \qquad
 \eta_{i_+}=:\frac{1}{\sqrt{2}}\eta^\prime_\iota~,
 \qquad
 \mathcal C^-_{\mu\nu\,i_+ j_+}=:\frac{\kappa}{4} T^-_{\mu\nu\,\iota\varsigma}~,
 \qquad \chi_{i_+}^\text{L}=:\sqrt{\frac{3}{8}}\kappa\chi_\iota^\prime~,\\
 a_\mu&=:\frac{\kappa}{2} \mathcal A_\mu~,
 \qquad
 i A_{\mu}^I\left(\Gamma_I\right)_{i_+}^{\ \ j_+} =: \frac{\kappa}{\sqrt{8}}V_{\mu\,\iota}^{\ \ \,\varsigma}~,
 \qquad
 \varphi=:\sqrt{\frac{3}{8}}\kappa\varphi^\prime~,\nonumber
\end{flalign}
where $\iota:=i_+$, $\varsigma:=j_+$, 
the expressions for the auxiliary fields are
\begin{align}
\begin{split}
 \omega_{\mu a b}&=\omega_{\mu a b}(e)
       -\frac{1}{4}\kappa^2\left(
    i{\bar\Psi}^{\iota}_{a}\gamma_{\mu}\Psi_{b \iota}
    +2i{\bar\Psi}_{\mu}^{\iota}\gamma_{[ a}\Psi_{ b] \iota}
   +\text{c.c.}\right)~,
  \\
  \varPhi_{\mu \iota}&=   
     -\frac{1}{2}i\Big(\gamma^{\nu\rho}\gamma_{\mu}-\frac{1}{3}\gamma_{\mu}\gamma^{\nu\rho}\Big)
      \Big(D_\nu\Psi_{\rho \iota} -\frac{\kappa}{16}\gamma\cdot T_{\iota\varsigma}^-\gamma_{\nu}\Psi_{\rho}^{\varsigma}\Big)
      +\frac{1}{2}\gamma_\mu\chi^\prime_\iota
      ~,
  \\
  \hat R^\prime_{\mu\nu\,\iota}(Q) &= 2 D_{[\mu}\Psi_{\nu]\iota}-i \gamma_{[\mu}\varPhi_{\nu]\iota}-\frac{\kappa}{8}\gamma\cdot T^-_{\iota\varsigma}\gamma_{[\mu}\Psi_{\nu]}^{\varsigma}~.
\end{split}
\end{align}
With the Fierz identity
\begin{align}
 v^iw^{}_j=
   \frac{1}{4}\,v^k w^{}_k\,\delta_j^{\ \,i} 
   +\frac{1}{4}\,v^k(\Gamma_m)_k^{\ \:l}w^{}_l\,(\Gamma_m)_j^{\ \,i}
   -\frac{1}{8}\,v^k(\Gamma_{mn})_k^{\ \:l}w^{}_l\,(\Gamma_{mn})_j^{\ \,i}
\end{align}
the transformations (\ref{eqn:delta-zeta-transformation}), (\ref{eqn:delta-eta-transformation}) 
to leading order in the fermions are
\begin{align}
\begin{split}
 \delta e_\mu^a &= - i \kappa \bar\zeta^{\prime\iota}\gamma^a\Psi_{\mu \iota} +\text{c.c.}~,\\
 \delta \Psi_{\mu \iota} &= 2\kappa^{-1}D_\mu\zeta^\prime_\iota-\frac{1}{8}\gamma\cdot T_{\iota\varsigma}^{-}\gamma_{\mu}{\zeta^{\prime}}^{\varsigma}-i\kappa^{-1}\gamma_{\mu}\eta^\prime_\iota~,\\
 \delta T_{ab\,\iota\varsigma}^- &= 8i\bar{\zeta^{\prime}}_{[\iota}\hat R^\prime_{ab\,\varsigma]}(Q)~,\\
 \delta V_{\mu\,\iota}^{\ \ \,\varsigma} &= \left(2\bar\zeta^{\prime \varsigma}\varPhi_{\mu\,\iota}-3\bar\zeta^{\prime\varsigma}\gamma_\mu\chi^\prime_\iota-2\bar\Psi_\mu^\varsigma\eta^\prime_\iota-\text{h.c.}\right)_\text{traceless}~,\\
 \delta \mathcal A_\mu &= -\frac{1}{2}i\bar\zeta^{\prime\iota}\varPhi_{\mu\,\iota}-\frac{3}{4}i\bar\zeta^{\prime\iota}\gamma_\mu\chi_\iota^\prime+\frac{1}{2}i\bar\Psi_\mu^\iota\eta^\prime_\iota+\text{c.c.}~,\\
 \delta \chi^\prime_\iota &= -\frac{1}{12}\gamma\cdot T_{\iota\varsigma}^-{\eta^\prime}^\varsigma
                             -i\varphi^\prime\zeta^\prime_\iota
                             +\frac{i}{12}\gamma\cdot\big(T_{\iota\varsigma}^-  \Dleftslash\big){\zeta^\prime}^\varsigma
                             -\frac{1}{3}\gamma\cdot R(\mathcal A)\,\zeta^\prime_\iota
                             -\frac{1}{6}i\gamma\cdot R(V)^{\ \,\varsigma}_{\iota}\,\zeta^\prime_\varsigma~,\\
 \delta\varphi^\prime &= \bar\zeta^{\prime \iota}\gamma^\mu\Big(D_\mu-\frac{\kappa}{2}\delta_{\zeta^\prime}(\Psi_{\mu}) -\frac{\kappa}{2}\delta_{\eta^\prime}(\varPhi_{\mu}) \big)\chi^\prime_\iota+\text{c.c.}~,
\end{split}
\end{align}
where
$iF_{\mu\nu}^I(\Gamma_I)_{i_+}^{\ \: j_+}=:\frac{\kappa}{\sqrt{8}}R_{\mu\nu}(V)_\iota^{\ \,\varsigma}$
and 
$f_{\mu\nu}=:\frac{\kappa}{2}R_{\mu\nu}(\mathcal A)$.
These are the results obtained in \cite{deWit:1979ug, deRoo:1980mm} 
in Euclidean signature
up to differences in the phase factors.

\bibliographystyle{JHEP-2.bst}
\bibliography{N4sugra}

\end{document}